\documentclass{aa}  
\usepackage{graphicx}
\usepackage{txfonts}
\usepackage{xspace}
\usepackage[colorlinks=true, linkcolor=blue, citecolor=blue, urlcolor=blue]{hyperref}
\usepackage[normalem]{ulem}
\usepackage{xcolor}
\usepackage{amssymb}
\usepackage{cancel}
\usepackage{wrapfig}

\def\gaia{\textit{Gaia}\xspace}
\def\kepler{\textit{Kepler}\xspace}

\begin{document} 
   \title{Variability classification of TESS targets in LOPS2, the first long-term pointing field of PLATO}
   \subtitle{Version\,1 of the public variability catalogue}

   \author{Mykyta~Kliapets
          \inst{1,}\inst{2}
          \and
          Pablo~Huijse
          \inst{1}
          \and
          Jeroen~Audenaert
          \inst{2}
          \and
          Andrew~Tkachenko
          \inst{1}
          \and
          Marek~Skarka
          \inst{3}
          \and
          Paul~F.~X.~Gregory
          \inst{2}
          \and
          Dominic~M.~Bowman
          \inst{4,}\inst{1}
          \and
          Simon~J.~Murphy
          \inst{5}
          \and
            Poojan~Agrawal
            \inst{1}
            \and
            József~M.~Benkő
            \inst{6}
            \and
            Hannah~Brinkman
            \inst{1}
            \and
            Nicholas~Jannsen
            \inst{7,}\inst{1}
            \and
            Yoshi~Nike~Emilia~Eschen
            \inst{8}
            \and
            Allison~Eto
            \inst{2}
            \and
            Dario~J.~Fritzewski
            \inst{1}
            \and
            Alex~Kemp
            \inst{1}
            \and
            Viktor~Khalack
            \inst{9}
            \and
            Gang~Li
            \inst{5}
            \and
            Ricardo Ochoa-Armenta
            \inst{1}
            \and
            Inês~Rolo
            \inst{10,}\inst{11}
            \and
            Nena~Scheller
            \inst{12}           
            \and
            Rose~S.~Stanley
            \inst{1,}\inst{13}
            \and
            Keegan~Thomson-Paressant
            \inst{4}              
            \and
            Emese~Plachy
            \inst{7}
            \and
            Vincent~Vanlaer
            \inst{1}
            \and
            Mathijs~Vanrespaille
            \inst{1}
            \and
            Jasmine~Vrancken
            \inst{1}
            \and
            Haotian~Wang
            \inst{1}
            \and
            Yian~Xia
            \inst{1}
          \and
          George~R.~Ricker
          \inst{2}
          \and
          Conny~Aerts
          \inst{1,}\inst{14,}\inst{15}
          }

   \institute{Institute of Astronomy, KU Leuven, Celestijnenlaan 200D, bus 2401, 3001 Leuven, Belgium\\ 
              \email{mykyta.kliapets@kuleuven.be, pablo.huijse@kuleuven.be}
         \and 
             MIT Kavli Institute for Astrophysics \& Space Research, Massachusetts Institute of Technology, Cambridge, MA, USA
         \and 
             Astronomical Institute of the Czech Academy of Sciences, Fričova 298, CZ-25165 Ondřejov, Czech Republic
        \and 
            School of Mathematics, Statistics and Physics, Newcastle University, Newcastle Upon Tyne, NE1 7RU, UK
        \and 
            Centre for Astrophysics, University of Southern Queensland, Toowoomba, QLD 4350, Australia
        \and 
            Konkoly Observatory, HUN-REN CSFK, 1121 Budapest, Konkoly Thege u. 15-17, Hungary
        \and 
            Isaac Newton Group of Telescopes, Apartado de correos 321, E-38700 Santa Cruz de La Palma, Canary Islands, Spain
        \and 
            Department of Physics, University of Warwick, Gibbet Hill Road, Coventry CV4 7AL, UK
        \and 
            Université de Moncton, 18 avenue Antonine-Maillet, Moncton, N.-B. E1A 3E9, Canada
        \and 
            Instituto de Astrofísica e Ciências do Espaço, Universidade do Porto, CAUP, Rua das Estrelas, 4150-762 Porto, Portugal
        \and 
            Departamento de Física e Astronomia, Faculdade de Ciências, Universidade do Porto, Rua do Campo Alegre 687, 4169-007 Porto, Portugal
        \and 
            Institute for Astro- and Particle Physics, University of Innsbruck, Technikerstraße 25, 6020 Innsbruck, Austria
        \and 
            Department of Physics and Astronomy, Vrije Universiteit Brussel, Pleinlaan 2, 1050 Brussels, Belgium
        \and 
            Department of Astrophysics, IMAPP, Radboud University Nijmegen, PO Box 9010, 6500 GL, Nijmegen, The Netherlands
        \and 
            Max Planck Institute for Astronomy, Koenigstuhl 17, 69117, Heidelberg, Germany
             }

   \date{Received TBA; accepted TBA}
 
 
  \abstract
   {The PLAnetary Transits and Oscillations of stars (PLATO) mission is expected to launch in January 2027. A total of 8\% of its data rate will be dedicated to complementary science targets selected from approved Guest Observer proposals.}
{We seek to provide an open-source catalogue of variable stars in PLATO's first long-term observing field, LOPS2. We want to use existing observations from the Transiting Exoplanet Survey Satellite (TESS), which has observed many stars in LOPS2.}
   {We classified 38 million calibrated aperture light curves from the TESS-Gaia Light Curve pipeline (TGLC, $G\lesssim17$) for 6 million unique sources in LOPS2 with two machine learning frameworks -- a deep neural network and a feature-based gradient-boosted decision-tree ensemble. We combined their predictions to create this first version of the LOPS2 variability catalogue, performed manual vetting of a sub-sample classified light curves, and a statistical analysis of the results to validate our methodology and to assess the variability properties and parameters of the stars in the catalogue.}
   {Our classification resulted in the identification of approximately 72\% of the light curves having dominant instrument- or pipeline-induced signal, with the remaining 28\% representing 3.6 million individual candidate variable stars, including pulsating, rotating, and eclipsing stars. Candidate pulsators exhibit varied behaviour in terms of their frequencies, amplitudes, rotation, and fundamental parameters. To ensure purity of the samples, filtering on colour, luminosity, the dominant frequency and its amplitude, and presence of close neighbours is helpful.}
   {We provide the first version of our PLATO LOPS2 variability catalogue to the community for further study and scrutiny. It is to date one of the largest catalogues of variable stars from an automated classification pipeline.}

\authorrunning{M. Kliapets et al.} 
\titlerunning{PLATO LOPS2 Variability Catalogue, Version1} 

\maketitle
%
\section{Introduction}

PLATO \citep{rauer2025plato} is the upcoming Medium\,3 mission of the European Space Agency's (ESA)
Cosmic Vision 2015–2025 program,
scheduled to launch  in January 2027. 
PLATO is designed to detect and characterise star-planet systems with Sun-like host stars. Achieving PLATO's core science objectives \citep[][]{cabrera2026assessment} involves determining both stellar and planetary parameters from PLATO's light curves for F5 to M dwarfs and subgiants \citep[see, e.g.\ ][]{nascimbeni2025plato,Prisinzano2026},
as well as performing follow-up ground-based observations 
\citep[see, e.g.\ ][]{Udry2024,deeg2024ground}.

PLATO's payload includes 24 cameras (N-CAMs) operating in white light and two fast cameras (F-CAMs) equipped with blue (505-700 nm) and red (665-1000 nm) filters \citep[][]{pertenais2021unique,rauer2025plato}. The satellite is scheduled to observe the first long-term pointing field, LOPS2 \citep[][]{nascimbeni2025plato}, for two years, followed by either another two years of observations of LOPS2, or a switch to another field, such as the one identified in the northern hemisphere, LOPN1 \citep[][]{nascimbeni2022plato}.
The N-CAMs will produce on-board-processed light curves and $6 \times 6$ pixel images, called imagettes, at various cadences \citep[][]{rauer2025plato}. 
In terms of covered sky-fields, the N-CAMs of PLATO are grouped into $n_{N-CAMs} \in \{6,12,18,24\}$ configurations
\citep[see][for illustrations]{rauer2025plato,jannsen2025mocka}.

The targets to be observed by PLATO to achieve its main science objectives 
are included in the PLATO Input Catalogue \citep[PIC,][]{montalto2021all}. They
include dwarf and sub-giant stars of spectral class F5 to K7 with $V \leq 11$ (P1 sample) and $V \leq 8.5$ (P2 sample; a sub-sample of P1), cool K to M dwarfs monitored during long pointings (P4 sample), and a statistical sample of F5 to K7 dwarfs and sub-giants (P5 sample).\footnote{We note that there was originally also a P3 sample, but this was dropped during the mission preparation.} In addition to these core science stars \citep[tPIC,][]{montalto2026plato,nascimbeni2026plato}, PLATO will also observe fine guidance stars \citep[fgPIC,][]{heller2026plato}, instrument calibration stars \citep[cPIC,][]{heller2026plato}, and science calibration and validation stars (scvPIC). The latter in particular are 
essential to calibrate and validate the PLATO pipelines and to improve the input physics of stellar models, notably to reach 10\% age accuracy for the host stars. The selection and role of the scvPIC stars for LOPS2 are detailed in \citet[][]{zwintz2026plato}.

Aside from the main aims of PLATO covered by its core science program, the mission also offers observing time to the worldwide community to perform any type of science that the instrument 
can help solve, provided that it is not yet covered in the core program. This so-called
PLATO Complementary Science (PLATO-CS) program
is allocated 8\% of PLATO's telemetry budget. The PLATO-CS programme 
will be filled through competitive, open Guest Observer (GO) calls by ESA, 
the first occurring on 7 April 2026.\footnote{\url{https://www.cosmos.esa.int/web/plato/getting-ready}}
Through GO proposals, the community will be able to request to observe targets in the predetermined PLATO field (LOPS2 for the first GO). 

Preparatory work in examples of PLATO-CS research covers 
distinct science cases. For stellar science it concerns pulsating stars across the Hertzsprung-Russell diagram (HRD), binary and multiple stars, magnetic fields of stars, rotational variability, mass loss of stars, stars with debris disks, and more. Stellar populations, including clusters, associations, and moving groups also form natural targets, given PLATO's large LOPS2 (2\,149 deg$^{2}$). Beyond these galactic topics, extragalactic topics are also possible for GO applications, notably to monitor compact binaries and the transient Universe in general. Some of the preparatory activities for PLATO-CS, showing that the instrument is well capable of global variability studies, were presented in 
\citet{Tkachenko2024,AertsTk2024,Audenaert2024,SouthworthMaxted2025,jannsen2024platosim,jannsen2025mocka}.

PLATO partially builds on the success of \kepler \citep[][]{borucki2010} and TESS \citep[][]{ricker2015transiting} in exoplanet hunting and stellar variability studies.
Unlike PLATO, TESS is an all-sky survey, which has observed more than $98\%$ of the sky, each star covered in at least one of its 27.4-d sectors, including complete coverage of LOPS2 \citep[][]{eschen2024viewing}.  

Our global scientific aim for this paper is twofold:
\begin{enumerate}
    \item to determine and interpret the variability content of LOPS2 by adopting a homogeneous analysis framework, capable of deducing the parameter space covered by objects in the field from their public TESS photometry; and 
    \item to provide the community with Version\,1 of our open-source catalogue of variable objects in LOPS2.\footnote{Our catalogue is made available on Zenodo via \url{TBA} [when accepted].}
\end{enumerate}

We achieve our aims by performing homogeneous automated
variability classifications using machine learning methods \citep[see][for a recent overview]{audenaert2025stellar}.
Even with a certain number of false-positive detections -- unavoidable for automatic classification methods -- our variability catalogue will substantially reduce time and work investments in selecting potential targets for PLATO-CS GO proposals. However, our open-access variability catalogue
can also be used beyond the needs of PLATO-CS. These include variability studies in TESS without subsequent use of PLATO data or integrations into multi-technique datasets used in foundation machine learning architectures such as the method presented in \citet[][]{angeloudi2024multimodal}.
We present our TESS dataset in Sect.\,\ref{sec:data} and our variability classifiers in Sect.\,\ref{sec:res}. 
The optimal ways to navigate our catalogue are discussed in Sect.\,\ref{sec:navig} and we summarise the astrophysical properties of the variables in LOPS2 in Sect.\,\ref{sec:param}. We discuss our global results in Sect.\,\ref{sec:disc} and summarize our main conclusions in Sect.\,\ref{sec:conc}.

\section{The dataset}\label{sec:data}

We describe our training set composition, including assembly and vetting, as well as the unlabelled data used to construct the variability catalogue.

\subsection{TESS-Gaia light curves}\label{sec:tglc}

Variability studies are typically performed using light curves, frequency (or period) spectra, or both. In this project, we made use of TESS time-series photometry. TESS data products used by the variable star community include the light curves created by the
TESS Science Processing Operations Center \citep[SPOC,][]{caldwell2020tess}, TESS Asteroseismology Science Operation Centre \citep[TASOC,][]{handberg2021tess,lund2021tess}, Quick Look Pipeline \citep[QLP,][]{huang2020photometry,huang2020photometry2,kunimoto2021quick,kunimoto2022quick}, and TESS-\gaia Light Curves \citep[TGLC,][]{han2023tess}. As announced by the TESS Science Office\footnote{\url{https://tess.mit.edu/qlp/}} at the Massachusetts Institute of Technology, where TESS data are processed, the TGLC method for contamination removal \citep[][]{han2023tess} will be implemented into the QLP pipeline \citep[][]{petitpas2026qlp}. Hence we used TGLC light curves. Previous studies revealed that they show good agreement in extracted dominant and secondary variability with QLP \citep[][]{kliapets2025automated}, as well as with \gaia \citep[][]{hey2024confronting}. TGLC data also work well for crowded fields \citep[][]{han2023tess}, which is crucial for anticipated cluster asteroseismology studies 
\citep[e.g.,][]{Bedding2023,li2024asteroseismology,fritzewski2024age,Fritzewski2026,wang2026}
 to be done in LOPS2.

Each calibrated aperture TGLC light curve was preprocessed in the same way as in \citet[][]{hey2024confronting} and \citet[][]{kliapets2025automated}, that is we applied both TESS and TGLC quality flags to remove low-quality observations. We then clipped outliers and subtracted a Gaussian-smoothed time series with a width of 100 $\sigma_{G}$  datapoints to remove some of the potential long-period (low-frequency) systematic trends. The differences with previous approaches is that our outlier clipping was done at 
5\,$\sigma_{out}$ for the upper limit and at 10\,$\sigma_{out}$ for the lower limit in order not to remove eclipses from the data. We note that some light curves still display systematics at certain frequencies 
(see Sect.\,\ref{sec:training}); the composition of our training set is designed with that in mind. Following preprocessing, light curves in the nominal and extended mission have $1300 \pm 150$ and $3670 \pm 270$ measurements, respectively. Noise budget for a sub-sample of light curves used in this study is shown in Figure\,\ref{fig:noise}; see also \citet[][]{sullivan2015transiting} and \citet[][]{kunimoto2022predicting}.

The previously described preprocessing steps are in addition to the steps already taken to detrend the calibrated apperture photometry \citep[][]{han2023tess}. The built-in calibrated aperture TGLC preprocessing relies on detrending with a 1-d window \citep[][]{han2023tess}, removing long-period variability for some light curves. In the preparation of the training set (Sect.\,\ref{sec:training}), we found that Cepheids with long periods \citep[][]{plachy2021tess} are adversely affected and the loss of their dominant frequency made them appear to have noisy light curves. Given that both the training set (Sect.\,\ref{sec:training}) and unlabelled TESS light curves (Sect.\,\ref{sec:unlab}) are subject to the same treatment, this is expected to limit the detectability of long-period Cepheids in the variability catalogue by increasing false negatives.

\subsection{The training set}\label{sec:training}

We used the training set from \citet[][]{gregory2026astrafier} as a baseline, which in turn was based on \citet[][]{audenaert2021tess}. Since it was constructed from QLP light curves, we cross-matched TESS input catalogue (TIC) identification numbers (IDs) with \gaia Data Release 3 (DR3) IDs to retrieve all available TGLC light curves available for bulk download from The Barbara A. Mikulski Archive for Space Telescopes (MAST).\footnote{\url{https://archive.stsci.edu/hlsp/tglc}}

Because QLP and TGLC light curves have different instrumental and sector-specific artefacts and can report different dominant variability due to differences in calibrations and/or preprocessing, we vetted the training set in a three-step process:

\begin{enumerate}
    \item High-level statistical vetting: analysing distributions of effective temperature ($T_{\mathrm{eff}}$) from \textit{Gaia} DR3 data \citep[][]{prusti2016gaia},
dominant variability $f_1$, and its amplitude $A_1$ \citep[][]{de2023gaia}, 
as well as stacked amplitude spectra per class \citep[e.g.,][]{li2020gravity,hey2024confronting}. Light curves associated with parameter values or behaviour not expected for a certain class were removed;
    \item Frequency systematic vetting: inspecting light curves where $f_1$ is one of the TESS or TGLC frequency systematics (discovered through manual light curve inspection) at \( \{0; 0.9; 0.92; 0.93; 1\}~\mathrm{d}^{-1} \) with a tolerance of 0.02 d$^{-1}$ — see \citet[][]{hey2024confronting} for a discussion of the 1~d$^{-1}$ systematic. These were manually reclassified as light curves dominated by instrumental effects or removed;
    \item Low-level light curve vetting: inspection of a sub-sample of the training set, particularly one nominal and one extended mission light curve per target. Light curves or frequency spectra displaying morphology not expected for a certain class were removed.
\end{enumerate}

The training set comprises seven variability classes. With the exception of the APERIODIC class, all variables are captured well on a time-scale of 27.4~d of a single TESS sector. The eighth INSTRUMENT class was already introduced by \citet[][]{gregory2026astrafier}. In this study, the latter class includes members with both TESS- and TGLC-specific systematics, whereas in \citet[][]{gregory2026astrafier} it included stars with TESS and QLP systematics. We refer the reader to \citet[][]{aerts2010asteroseismology} and \citet[][]{audenaert2021tess} for extensive discussions of variable star classes and to \citet[][]{aerts2024asteroseismic} and \citet[][]{bowman2026asteroseismology} for reviews of recent discoveries of selected classes of pulsating stars, aside from the INSTRUMENT class discussed in \citet[][]{gregory2026astrafier}. 
Here we limit ourselves to a brief description of the eight classes for the reader's convenience:
\begin{enumerate}
    \item APERIODIC: a class of aperiodic variables, whose full variability cycle is not captured by a single TESS sector, such as Mira or long-period rotational variables. In the data we used, this class is characterised by low variability amplitudes, with most of the power excess concentrated at low frequencies;
    \item CONTACT\_ROT: a class of contact binaries and stars with surface inhomogeneities, such as Ap/Bp stars. This class typically has a sinusoidal-looking light curve and high rotational peaks with (sub-)harmonics in the frequency space;
    \item DSCT\_BCEP: a class of intermediate- ($\delta$\,Sct stars) and high-mass ($\beta$\,Cep stars) radial and non-radial pulsators exhibiting pressure (p) mode pulsations. The pulsators in this class typically have frequencies above 5~d$^{-1}$. The class includes hybrid pulsators also experiencing lower-frequency gravity (g) modes. Following the Initial Mass Function (IMF), the class population is skewed towards the intermediate-mass regime;
    \item ECLIPSE: a class with eclipsing binaries or transits with at least one occurrence visible in a single-sector TESS light curve;
    \item GDOR\_SPB: a class of intermediate-mass F-type ($\gamma$ Dor stars) or Slowly Pulsating B-type (SPB) stars experiencing high-order non-radial g-mode pulsations or Rossby (r) modes. The pulsators in this class typically have frequencies below 5~d$^{-1}$, The class includes hybrid p- and g-mode pulsators and is skewed towards the F-type regime following the IMF;
    \item RRLYR\_CEPH: a class of high-amplitude classical radial pulsators (RR Lyrae and Cepheids), characterised by a particular periodic and mostly non-sinusoidal light curve shape. Following the IMF, this class is skewed towards RR Lyrae stars and only includes short-period Cepheids (see Sect. \ref{sec:tglc});
    \item SOLARLIKE: a class of dwarfs, subgiants, or red giants displaying stochastic and/or rotational variability at low frequency, superposed with high radial-order low-amplitude p-modes at high frequency. The members in this class reveal a noisy floor in most of the frequency spectrum;
    \item INSTRUMENT: a joint class of stars dominated by TESS- or TGLC-specific instrumental systematics, as well as stars without variability. These are characterised by noisy light curves and frequency spectra with typically only instrumental peaks passing false alarm probability or signal-to-noise ratio criteria.
\end{enumerate}
Compared to \citet[][]{gregory2026astrafier}, we enriched the class of RRLYR\_CEPH with more light curves of Cepheids from \citet[][]{ripepi2023gaia} to compensate for the large imbalance between Cepheids and RR Lyrae stars. 

Overall, this resulted in a training set containing 26\,521 light curves, of which 19\,358 and 7\,163 from the nominal and extended TESS missions, respectively (Table~\ref{tab:training_set}). 
A table containing TIC and \gaia DR3 IDs, coordinates, as well as value-added information for each light curve is available on Zenodo. Note that we used all available TGLC light curves for the training set, regardless of whether they fall into the PLATO field of view (most do not); we revisit this in Sect.\,\ref{sec:disc}.

\begin{table*}[ht!]
\centering
\caption{Number of light curves in the training set from the nominal (NM, first row) and first extended (EM, second row) mission.}
\label{tab:training_set}
\tabcolsep=2pt
\begin{tabular}{lcccccccc}
\hline
\hline
Class & 
APERIODIC & CONTACT\_ROT & DSCT\_BCEP &
ECLIPSE & GDOR\_SPB & INSTRUMENT & RRLYR\_CEPH & SOLARLIKE \\
\hline
NM &
836 & 2\,371 & 5\,104 & 759 & 
3\,354 & 4\,019 & 854 & 1\,888 \\
EM &
685 & 960 & 1430 & 398 &
852 & 1\,672 & 320 & 686 \\
\hline
\end{tabular}
\end{table*}

\subsection{LOPS2 stars observed by TESS}\label{sec:unlab}

\begin{figure}
    \centering
    \resizebox{\hsize}{!}{\includegraphics{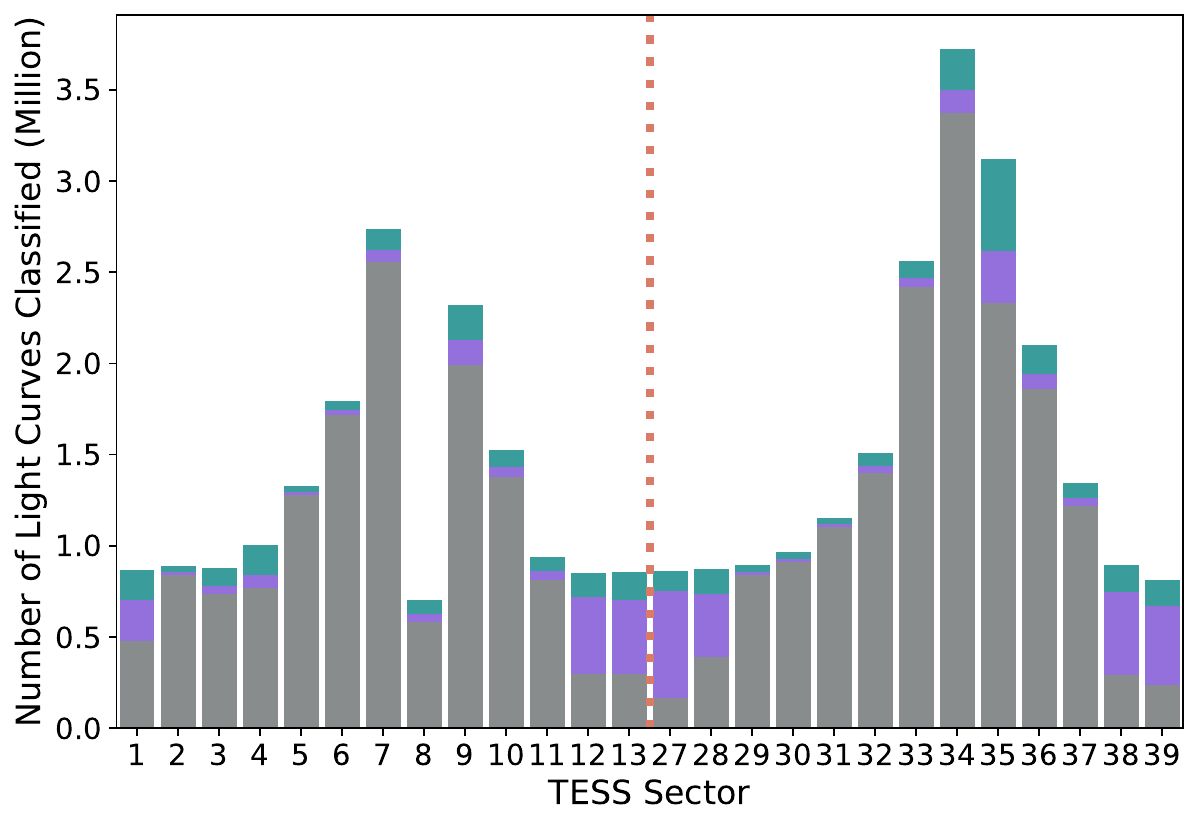}}
    \caption{Number of light curves for stars in LOPS2 classified per TESS sector. The entire bin height indicates all light curves for a given sector, where those with a dominant instrumental frequency $f_1$ are indicated in purple and those with a second or third instrumental frequency (without double counting) in teal. The burgundy dotted line delineates classified nominal and extended mission data.}
    \label{fig:systematic}
\end{figure}

We classified all stars observed by TESS occurring in LOPS2, as queried from the LOPS2 catalogue in \citet[][]{jannsen2025mocka}. This includes nominal mission data from sectors 1-13 and the first extended mission data from sectors 27-39. In total, we classified 16\,694\,681 and 20\,812\,306 light curves of these two groups of sectors, respectively. Some light curves have an empty flux array or negative calibrated flux values. Hence, we excluded these targets from the classification framework. In Fig.\,\ref{fig:systematic}, we show the distribution of the number of available light curves per TESS sector. Each light curve was classified per sector separately to avoid stitching different TESS-sector data with inconsistent quality. Therefore, light curves of the same object observed in different TESS sectors may receive different classifications; we revisit this in Appendix\,\ref{appendix:consist}.

Some TESS sectors are more affected by instrumental and/or pipeline systematics than others. Figure\,\ref{fig:systematic} also shows the distribution of the stars with dominant or secondary instrumental frequencies at \( \{0; 0.9; 0.92; 0.93; 1\}~\mathrm{d}^{-1} \) with a tolerance of 0.02~d$^{-1}$ (without double-counting). Sectors 27 and 35 are the most affected in absolute numbers in the nominal and extended mission, respectively, while sectors 27 and 39 are the most affected relative to the total light curve number. 
For most sectors that display an instrumental frequency, it is often the dominant frequency $f_1$.
The most extreme exception is sector 35, potentially containing many stars with real variability yet some instrumental signal as second ($f_2$) or third ($f_3$) frequency. This suggests that these signals are pipeline-, rather than TESS-specific signals. 
Note that Fig.\,\ref{fig:systematic} should not be interpreted as a distribution of a fraction of light curves classified as instrumental (INSTRUMENT) by our pipeline (Sect.\,\ref{sec:arch}) because other factors in addition to the dominant, secondary, and tertiary frequencies contribute to the classification. A fraction of light curves predicted for each class, including INSTRUMENT, as a function of \gaia $G$ magnitude is shown in Fig.\,\ref{fig:frac}.

We performed a single classification of all light curves, binning the time stamps of the extended-mission data (i.e. 10~min) to the cadence of the nominal mission (i.e. 30~min). This is because our training set does not contain enough light curves from the extended mission to warrant two separate classifications. The effect of downsampling the cadence is discussed in Sect.\,\ref{sec:arch}.

\section{The classifiers}\label{sec:res}

We now describe our machine learning models, the final ensemble approach, and its evaluation.

\subsection{Individual classifiers}

\subsubsection{Deep neural network light curve classifier}\label{sec:arch}

Following up on \citet[][]{gregory2026astrafier}, we used their publicly available ASTRAFier architecture as our first individual classifier, which they used on TESS QLP light curves for the \kepler field of view (TESS sectors 14, 15, and 26). We refer the reader to \citet[][]{gregory2026astrafier} for an extensive description of the classifier and discussion of its performance. Here we only say that ASTRAFier is a deep learning model that embeds light curves into a high-dimensional space, and then processes it through repeated blocks of bidirectional long short-term memory, convolutional neural networks, and transformers, to capture long- and short term trends and the overall light curve morphology, respectively.

\begin{figure}
    \centering
    \resizebox{\hsize}{!}{\includegraphics{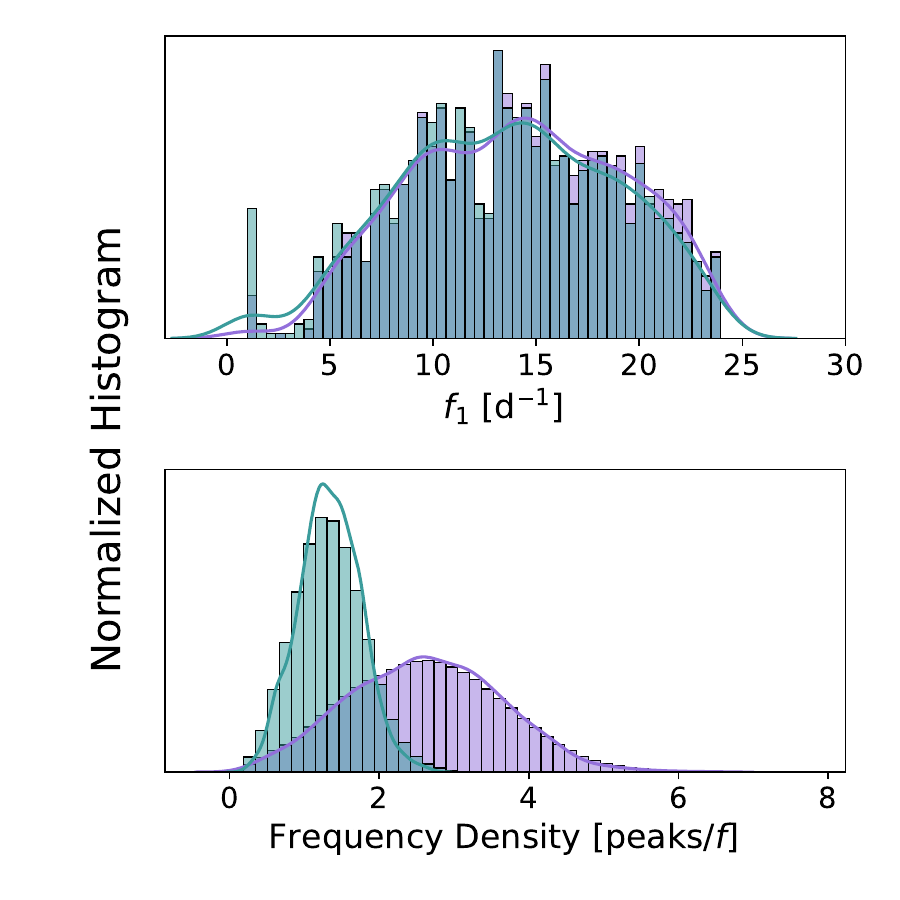}}
    \caption{Normalised distributions of the dominant frequency $f_1$ (top) and frequency density of the entire frequency space (bottom) before (purple) and after (teal) downsampling the cadence with a factor 3 for extended-mission p-mode pulsators. Histograms are plotted from 1\,000 samples for each candidate frequency within the uncertainty range. Kernel Density Estimators are plotted as full lines and were computed directly from the point estimates.}
    \label{fig:downsampling}
\end{figure}

We used the architecture from \citet[][]{gregory2026astrafier} as-is, with the exception that its preprocessing pipeline was made for nominal-mission QLP light curves only. We therefore included downsampling of extended-mission data to a 30-min cadence and an option to switch from QLP to TGLC light curves, as the two employ different quality flags. Compared to \citet[][]{gregory2026astrafier}, we also applied less strict preprocessing to these light curves, adopting a Gaussian smoothing of $\sigma_{G(TGLC)}=100$ instead of $\sigma_{G(QLP)}=61$. Downsampling 10-min cadence data to a 30-min cadence may influence mode detection in the high frequency range \citep[see][]{bowman2017}. The class most affected by this are the p-mode pulsators\footnote{To avoid confusion, here and further p-mode pulsator means $\delta$ Sct and $\beta$ Cep stars. Radial p-mode pulsators, RR Lyrae and Cepheids, are referred to as (classical) radial pulsators in this paper.} but this is one of the best retrieved classes (see Section \ref{sec:cm}). Indeed, in Fig.~\ref{fig:downsampling} we show the worst-case impact of binning extended-mission data for this class using confirmed p-mode pulsators from the training set. The top panel shows the distribution of the dominant mode before (purple) and after (teal) binning up to the Nyquist limit of binned data, while the bottom panel shows the frequency density (that is, the number of significant peaks per unit frequency) of the entire frequency space up to the Nyquist limit of binned data. A large shift in the density distribution (bottom panel) shows that some frequency information is clearly lost. However, the dominant mode (top panel) is not significantly affected. A shift in the density distributions can be explained by the binned time series not capturing signal at higher frequencies mirrored around the Nyquist limit due to the decrease in sampling frequency \citep[][]{hey2021search}, or by the effect on real p modes. The increase in the peak in the 1~d$^{-1}$ bin suggests that the dominant signal is of instrumental or pipeline nature but the light curve still exhibits real variability. We stress that other classes are affected less by the downsampling, which altogether justifies the use of downsampled data.

We trained our model on a single NVIDIA A2 GPU (16 GB VRAM) using mixed-precision training. Due to memory constraints, we used a batch size of 16 with gradient accumulation to obtain an effective batch size of 256.\footnote{See the Quick Start section of the README file: \url{https://github.com/jeraud/TESS-Transformer/tree/main}.} On more powerful hardware, one could increase the batch size and disable gradient accumulation. Training on the NVIDIA A2 GPU took us approximately 12 hours. Predicting on a light curve batch of 20\,000 light curves with 20 forward passes on one NVIDIA A2 GPU takes approximately 6~min (including data loading), each transformer iteration taking 0.47\,s.

\subsubsection{Feature-based boosted trees classifier}

For the second individual classifier, we used eXtreme Gradient Boosting \citep[XGBoost,][]{chen2015xgboost,chen2019package}, which is an ensemble by itself.\footnote{\url{https://platocs.miraheze.org/wiki/WP\_160\_000/Catalogue\_feature-based}.} XGBoost is an implementation of the gradient boosting algorithm \citep[][]{friedman2001greedy}, sequentially learning an ensemble of weak learners (decision trees), where each new tree focuses on correcting the errors made by previous ones. Therefore, its main difference from the random forest classifier \citep[][]{breiman2001random,parmar2018review}, is that the latter learns trees in parallel, while an XGBoost does it sequentially. XGBoost optimises an objective function that combines a classification loss with a complexity penalty, which helps to prevent overfitting \citep[][]{chen2015xgboost,chen2019package}.

Unlike a neural network, XGBoost is a feature-based classifier. This means it relies on a set of engineered features derived from the time and frequency domains, which were designed to distinguish objects belonging to different variability classes. We refer the reader to \citet[][]{cabral2018fats} for a general overview, and to \citet[][]{audenaert2021tess}, \citet[][]{barbara2022classifying}, \citet[][]{hey2024confronting}, and \citet[][]{ranaivomanana2025variability} for concrete examples of feature-based machine learning in astronomy. 

In our case we used 37 features from four different categories:
\begin{enumerate}
    \item Features characterising marginal flux distribution, including the mean, standard deviation, skewness, and kurtosis of the flux, as well as their robust or quantile-based counterparts and flux percentile ratios from \citet[][]{richards2011machine};
    \item Features characterising the autocorrelation function, including autocorrelation at lag one, first autocorrelation peak value, first autocorrelation peak lag, second autocorrelation peak value, second autocorrelation peak lag, and the first lag where the autocorrelation drops below 0.5;
    \item Features characterising the periodogram, including the highest peak and its power, amplitude ratios between the amplitude of the first three harmonics of the dominant frequency and the amplitude of the dominant peak, as well as the second and third highest peaks relative to the dominant peak, the amplitude ratio between the one of the peak at half the dominant frequency and the one of the dominant frequency, the entropy, and the mean of the amplitudes; 
    \item Features characterising the periodic morphology, including amplitudes and phases, as defined in \citet[][]{debosscher2007automated}, and the coefficient of determination.
\end{enumerate}

Our model learned 800 weak learners with a maximum depth of four, learning rate of 0.1, data sub-sampling of 0.6, and L2/L1 regularisation coefficients of 10 and 1, respectively. This was determined by performing a five-fold cross-validation splitting the training set by source ID, using the same split as used by the deep learning-based classifier. Training took approximately eight minutes, and feature computation (including data loading) and predicting on a light curve batch of 20\,000 takes approximately 19\,s using Intel(R) Xeon(R) CPU E5-2680 v3 @ 2.50GHz.

We compared the predictions on the training set and a random sub-sample of the unlabelled data for the neural network and the XGBoost, and found that the deep learning architecture overpredicts g-mode pulsators compared to the XGBoost model, labelling CONTACT\_ROT and SOLARLIKE stars as GDOR\_SPB more often than the feature-based one. 
For the CONTACT\_ROT case, this might be because of frequencies and amplitudes of (sub-)harmonics of rotational peaks, to which a feature-based classifier has direct access to specifically differentiate rotational variables and contact binaries from other variable classes. The confusion between 
SOLARLIKE and GDOR\_SPB can be explained by the neural network receiving normalised fluxes. Relying on the morphology only, which for these two classes can be similar and difficult to differentiate without explicit amplitude information as used by the XGBoost, may cause confusion for the deep learning model.

Additionally, we found that the deep learning model predicts the INSTRUMENT class less often than the XGBoost does. One potential explanation is that this class is not representative enough of all possible instrumental effects in a balanced way due to it coming from a limited selection of TESS sectors. We revisit this issue in Sect.\,\ref{sec:disc}.

\subsection{Ensemble model}\label{sec:ens}

Both the deep-learning-based and the feature-based classifier output pseudo-probabilities (scores) for each of the eight (variable) classes.\footnote{The pseudo-probabilities can be calibrated into probabilities using temperature scaling. However, we opted not to do this and instead use the scores themselves.} We combined the predictions from both classifiers in an ensemble, following \citet[][]{dong2020survey}. In that work,  probabilities of several classifiers are averaged. In our case, we average across only two classifiers. This simple approach to ensemble learning is known to improve accuracy and robustness of predictions, while reducing the uncertainty on the results \citep[][]{lakshminarayanan2017simple}. Such an approach strengthens high-confidence predictions made by both classifiers, such as when both agree on the highest assigned label.
This somewhat reduces the need for post-processing, allowing for a purer sample even when filtered by pseudo-probabilities only (see Sect.\,\ref{sec:probas}). 

\begin{figure}
    \centering
    \resizebox{\hsize}{!}{\includegraphics{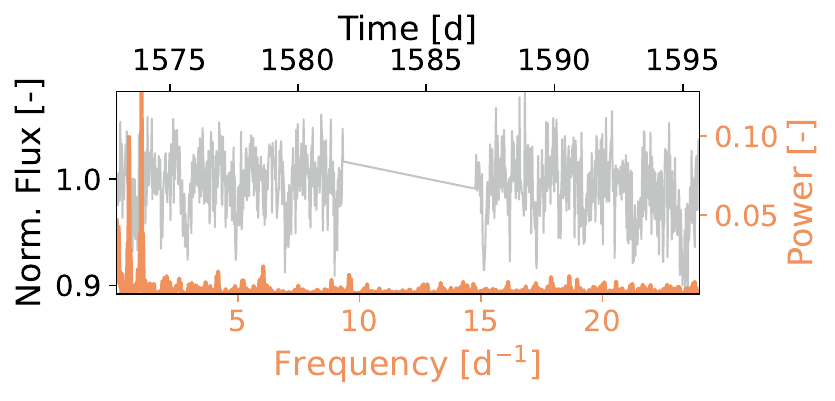}}
    \resizebox{\hsize}{!}{\includegraphics{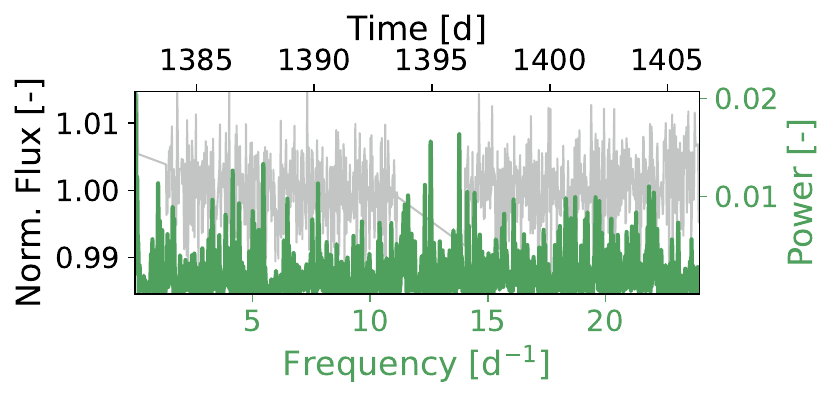}}
    \caption{Light curves (grey) and dimensionless Lomb Scargle periodograms as defined in 
    \citet{vanderplas2018understanding} overplotted in orange for \gaia DR3 5317171818865844864 predicted as CONTACT\_ROT by the neural network (0.85) and INSTRUMENT by XGBoost (0.55)
    from Sector 10 (upper panel) and in green for \gaia DR3 4775678010208086016 predicted as DSCT\_BCEP by the neural network (0.89) and INSTRUMENT by XGBoost (0.90) from Sector\,3. 
    The final prediction after averaging the scores is 0.6 of being CONTACT\_ROT (upper) and 0.72 of being INSTRUMENT (lower panel).}
    \label{fig:ens_err}
\end{figure}

When the classifiers disagree on the high-probability assignment, such as when one predicts class A with a high confidence, while another predicts class B, this either potentially excludes such an object from being selected when a particular score threshold is applied (when one classifier was right) or directly increases purity of the sample (when both are wrong). Examples of one model correcting another, resulting in a correct prediction, are shown in Fig.~\ref{fig:ens_err}, where the neural network is correct for the upper panel
and XGBoost is correct for the lower panel. In this plot (and further Lomb Scargle plots), the right y-axis is the dimensionless Lomb-Scargle power as defined in \citet[][]{vanderplas2018understanding}, relying on the standard Lomb-Scargle normalisation of \textit{astropy} \citep[][]{robitaille2013astropy,astropy2022astropy}.
We note that different models being inclined to favour different outcomes is a natural consequence of different inductive biases, which are necessary requirements for learning \citep[][]{mitchell1980need}.

The final content of the variability catalogue consists of eight scores for each of the eight classes of the ensemble, analogous to what was delivered by \citet[][]{audenaert2021tess}. The benefit of a (pseudo-)probabilistic output instead of a categorical class assignment is that it: (i) allows for flexible thresholding depending on the nature of a follow-up study and required sample sizes and purity; and (ii) enables flagging of hybrid (variable) classes, such as oscillators with both p and g modes, pulsators with rotational modulation, genuine variable stars with instrumental systematics, etc. However, split (pseudo-)probabilities can also occur as a result from the ensemble approach used rather than representing classes resulting from astrophysical properties.

\begin{figure}
    \centering
    \resizebox{\hsize}{!}{\includegraphics{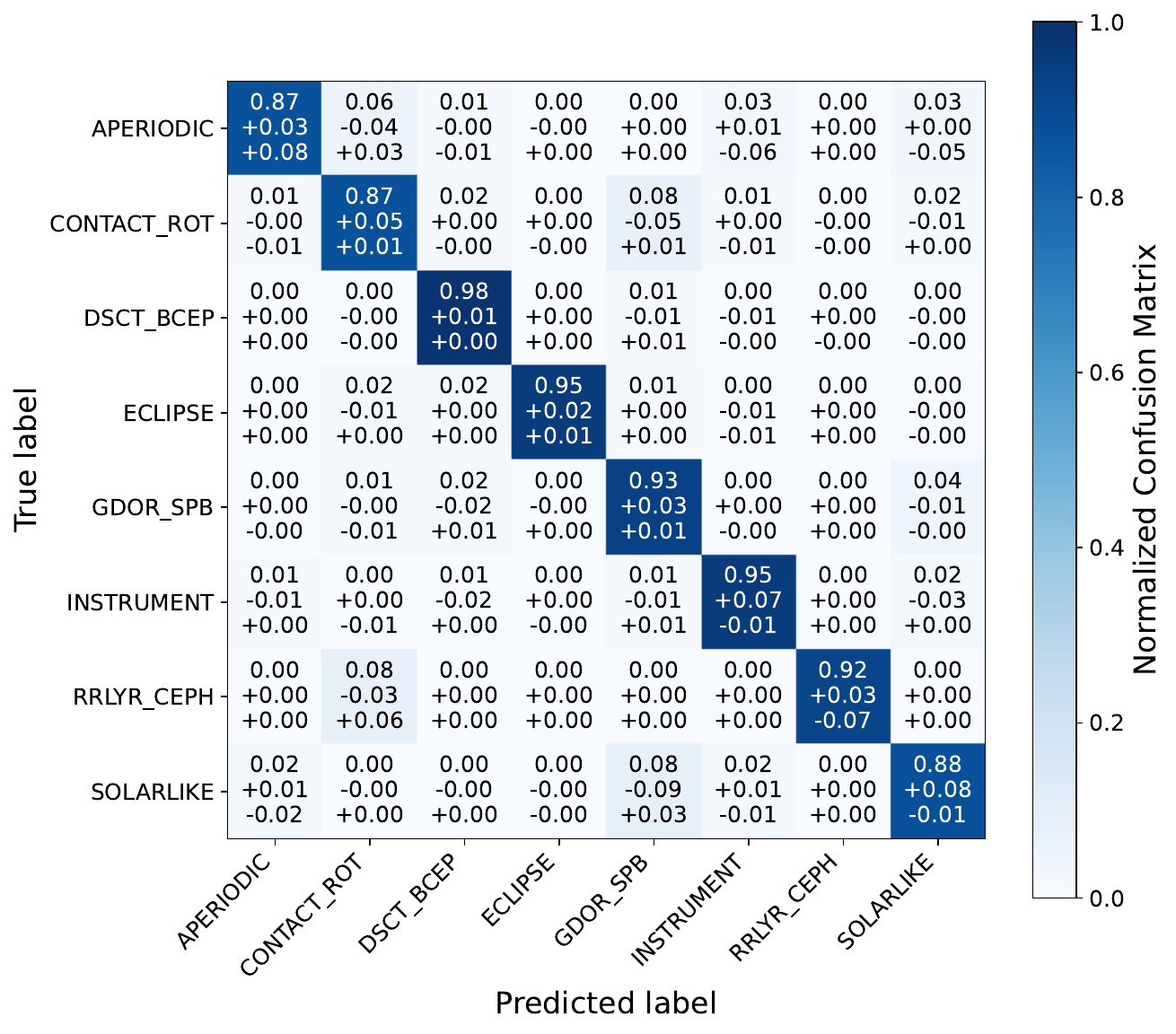}}
    \caption{The ensemble confusion matrix normalised by row. The first number in each cell is a fraction of the true label that is retrieved. The second and third rows are changes from using just the deep learning classifier and XGBoost, respectively. A positive delta (second and third row) is better on the main diagonal, and a negative delta is better off the main diagonal.}
    \label{fig:cm}
\end{figure}

\subsection{Classifier evaluation}\label{sec:cm}

We evaluate our ensemble on the hold-out set -- part of the training set put aside to test the classification performance -- comprising 20\% of our entire training dataset. Table\,\ref{tab:metr} shows the macro (all classes contributing equally) and weighted (classes weighted by the number of light curves in the hold-out set) precision (fraction of predicted positives that are actually positive), recall (fraction of positives found), and f-1 scores (harmonic mean of the former two). Each of these metrics is an improvement over each of the two individual models.

\begin{table*}[ht]
\centering
\caption{Precision, recall, and f-1 score for the ensemble. Values in parentheses are changes from using just the deep learning classifier and XGBoost, respectively.}
\begin{tabular}{lccc}
\hline
\hline
Average & Precision & Recall & f-1 score \\
\hline
Macro
& 0.932 ( +0.035 / +0.011 )
& 0.918 ( +0.040 / +0.005 )
& 0.925 ( +0.038 / +0.008 ) \\

Weighted
& 0.931 ( +0.037 / +0.005 )
& 0.930 ( +0.040 / +0.004 )
& 0.930 ( +0.040 / +0.005 ) \\
\hline
\end{tabular}
\label{tab:metr}
\end{table*}

A confusion matrix is shown in Fig.\,\ref{fig:cm} for both retrieved objects and changes from using just one of the two models. APERIODIC, CONTACT\_ROT, and SOLARLIKE are the most difficult classes for the model to predict, with the former and latter of these classes potentially due to challenges in distinguishing them in a single TESS sector. By combining the classifiers, both models predominantly gain recall, as can be seen by mostly positive deltas on the main diagonal of Fig.\,\ref{fig:cm}. One notable exception is the loss of 7\% for XGBoost in distinguishing radial pulsators, which the neural network confuses more with rotational variables than the feature-based classifier. In Appendices \ref{appendix:downs} and \ref{appendix:consist} we also investigated consistency of predictions on downsampled light curves for each class and across TESS sectors, respectively.

\section{Navigating the catalogue}\label{sec:navig}

\subsection{Thresholding}\label{sec:probas}

In addition to scores for each of the eight classes in the classification, we provide value-added information including the effective temperature ($T_{\mathrm{eff}}$), luminosity $L$, dominant variability frequency $f_1$ and its amplitude $A_1$ from sinusoidal fitting, etc. Additional information, particularly from \gaia, can be queried using \gaia DR3 IDs, provided in the catalogue.

It is ultimately up to the user to threshold predictions for a certain class. At the very minimum, we suggest to filter for the highest best class score (\texttt{max\_prob}) above 0.5. We stress that these scores do not represent probabilities in the usual meaning. Instead, they should be treated as similarities scores with respect to the (limited) training set used for both models. Increasing the \texttt{max\_prob} threshold generally leads to purer samples, although misclassifications cannot be excluded even at near-unity scores.

Thresholding by the dominant variability $f_1$ might be beneficial for classes where such cuts are unambiguous, such as for eclipsing binaries where the dominant frequency is rarely found above 4~d$^{-1}$. Similarly, aperiodic (long-period) variables could be thresholded at approximately 0.5~d$^{-1}$ and solar-like oscillators around 1.5~d$^{-1}$. For other classes, the situation becomes less clear, as the domains of dominant frequencies for some classes can overlap. For example, putting a threshold for (candidate) p-mode pulsators at 5~d$^{-1}$ might result in the inclusion of short-period contact binaries. Shifting it to frequencies above 4~d$^{-1}$ may create confusion with fast-rotating prograde g-mode pulsators as happened for the case of the SPB pulsator HD\,43317 -- see \citet[][]{Papics2012} versus \citet[][]{Buysschaert2018}.

Thresholds for $A_1$ could be put at ppt levels typically expected to be observable by TESS, such as 1\,ppt for main-sequence p- and g-mode pulsators of intermediate mass and rotational variables, while 0.1\,ppt for solar-like oscillators. Cuts in $T_{\mathrm{eff}}$ and $L$ should be applied with caution as \gaia temperatures and luminosities could be inaccurate (see Sect.\,\ref{sec:hrd}). As a final note, we encourage users to check nearest neighbours of stars in the catalogue, as bright sources can contaminate light curves of other stars due to the large TESS pixel size. An example of a tool one can use to remedy this is \texttt{TESS\_localize} \citep[][]{higgins2023localizing}, as demonstrated by \citet[][]{pedersen2023contamination}.

To minimise contaminants of different origin, we suggest the following workflow:

\begin{enumerate}
    \item Check pulsation behaviour in terms of frequencies and amplitudes and evaluate with respect to the expected behaviour/distribution for a given class \citep[][]{aerts2010asteroseismology,bowman2026asteroseismology};
    \item Check whether these behaviour and predictions are consistent across TESS sectors;
    \item Check whether a star falls within the expected position on the colour-magnitude diagram; 
    \item Check whether a star has close neighbours in equatorial coordinates exhibiting near-identical frequencies.
\end{enumerate}

\subsection{Representative failure cases}

\begin{figure}
    \centering
    \resizebox{\hsize}{!}{\includegraphics{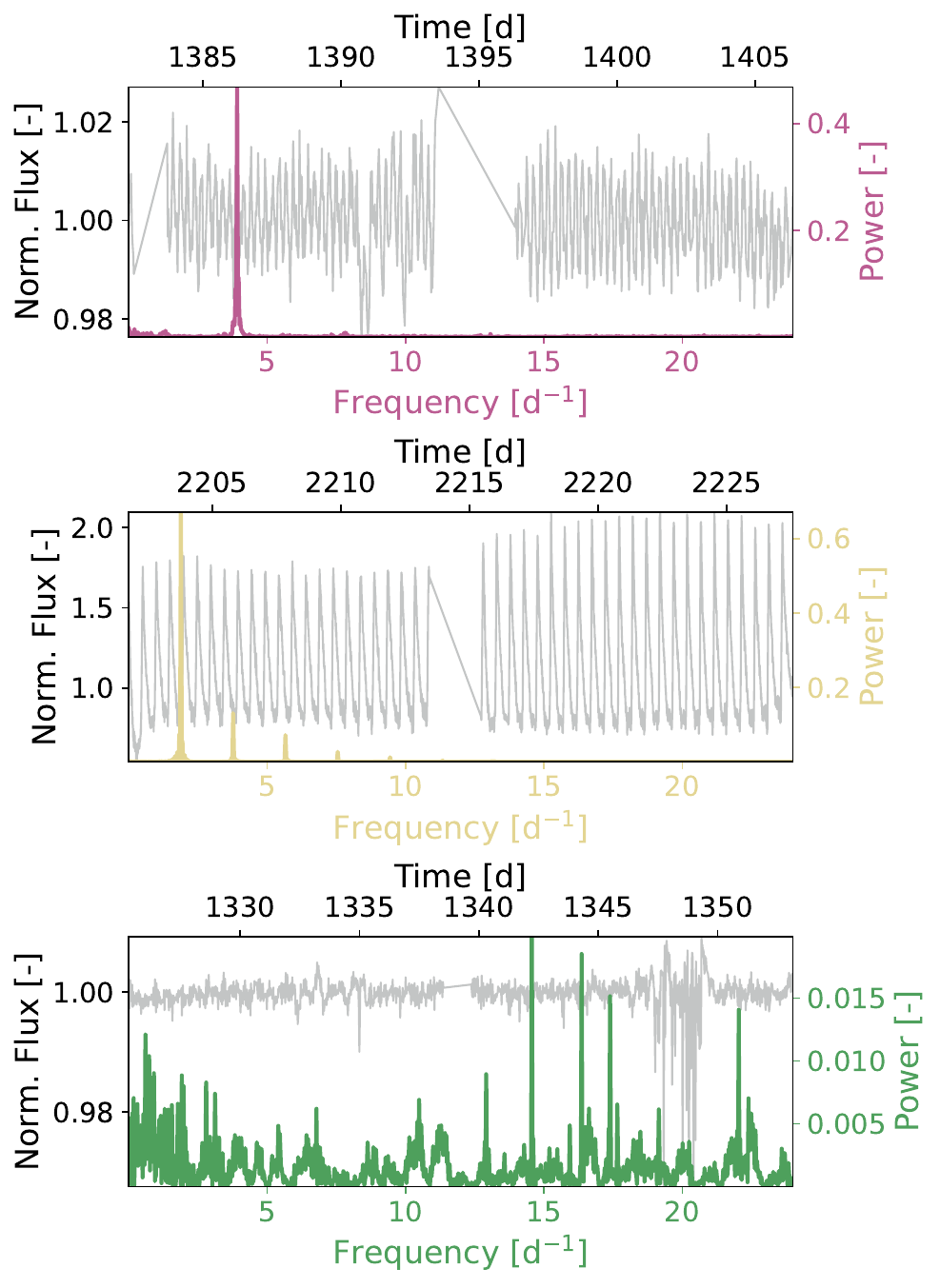}}
    \caption{Same as Fig.\,\protect\ref{fig:ens_err} for \gaia DR3 5486221491101551360 (sector 3, top), 2924953031585344896 (sector 33, middle), and 5290721760974905088 (sector 1, bottom), classified as GDOR\_SPB, RRLYR\_CEPH, and DSCT\_BCEP, respectively.}
    \label{fig:errors}
\end{figure}

Comprehensive error analysis of an automated variability classifier relies on manual inspection of vast amounts of data. We therefore show three hand-picked examples of representative failure cases in Fig.\,\ref{fig:errors} occurring for different reasons: variability-related, global parameter-related, and instrument-related. We note that only the first example represents an error in a machine learning context, while the latter two are relevant in the context of selecting candidate variables for follow-up studies or PLATO GO calls.

The top panel of Fig.~\ref{fig:errors} shows a light curve classified as GDOR\_SPB with a final score of 0.59 (the highest scores are 0.90 for being GDOR\_SPB by the neural network and 0.52 for being CONTACT\_ROT by XGBoost), which is a misclassified rotational variable. Automatic identification of g-mode pulsators, particularly their discrimination from rotational variables, is a hard learning problem, especially in this case where there is no (sub-)harmonic signal that would make it easier to separate the two. In this case the confusion is caused by a high score assigned by the neural network, which is more likely to assign higher pseudo-probabilities than XGBoost. This example demonstrates that thresholding on pseudo-probabilities alone, especially at low values, might not be sufficient to ensure purity of the selection.

The middle panel of Fig.~\ref{fig:errors} shows an example of a light curve correctly classified as RRLYR\_CEPH with a final score of 0.99 (with scores of 0.99 and 0.98 for the neural network and XGBoost, respectively). However, this light curve does not belong to an RR Lyrae star, since the target has a temperature (colour) and absolute magnitude inconsistent for this class of radial pulsators \citep[][]{plachy2021rr}. The reason for this is blending due to the large TESS pixels, an issue already previously reported for these stars \citep[][]{plachy2020rr}. This example demonstrates the importance of checking if a star has close neighbours or background contamination in a full frame image and evaluating its fundamental parameters. A full frame image for this source is shown in Fig.~\ref{fig:a3}. 

Finally, the bottom panel of Fig.~\ref{fig:errors} shows a light curve classified as a DSCT\_BCEP star with a final score of 0.99 (with scores of 1.00 and 0.99 for the neural network and XGBoost, respectively). While this classification might be correct (and therefore is not an error from a machine learning perspective), the light curve contains a strong instrumental effect common for TESS sector 1, resulting in a noisy floor in the frequency space and low power attributed to p-mode variability. This effect makes the light curve challenging for asteroseismic studies and may not be a convincing target for a GO proposal unless the data points associated with the systematics can be removed. This final example reinforces the need for detailed manual inspection even after candidate filtering (Sect.\,\ref{sec:probas}).

\section{Astrophysical properties}\label{sec:param}

We now provide a high-level overview of the fundamental and variability properties of the classified unlabeled set. The purpose of this study is to provide a synthesis of the variability of the stars in LOPS2 from TESS data and
basic information in the variability catalogue with respect to the parameter space for particular classes.

\subsection{Pulsation properties}\label{sec:puls}

An alternative way to evaluate performance of a variability classifier, in addition to the machine learning approaches and manual vetting described in Sect.\,\ref{sec:res}, is to compare the distribution of astrophysical properties with that of the training set. We first do so for the intrinsic variability due to pulsations, given that the TESS data is particularly powerful on this front.

\begin{figure*}[t!]
    \centering
    \includegraphics[width=18cm]{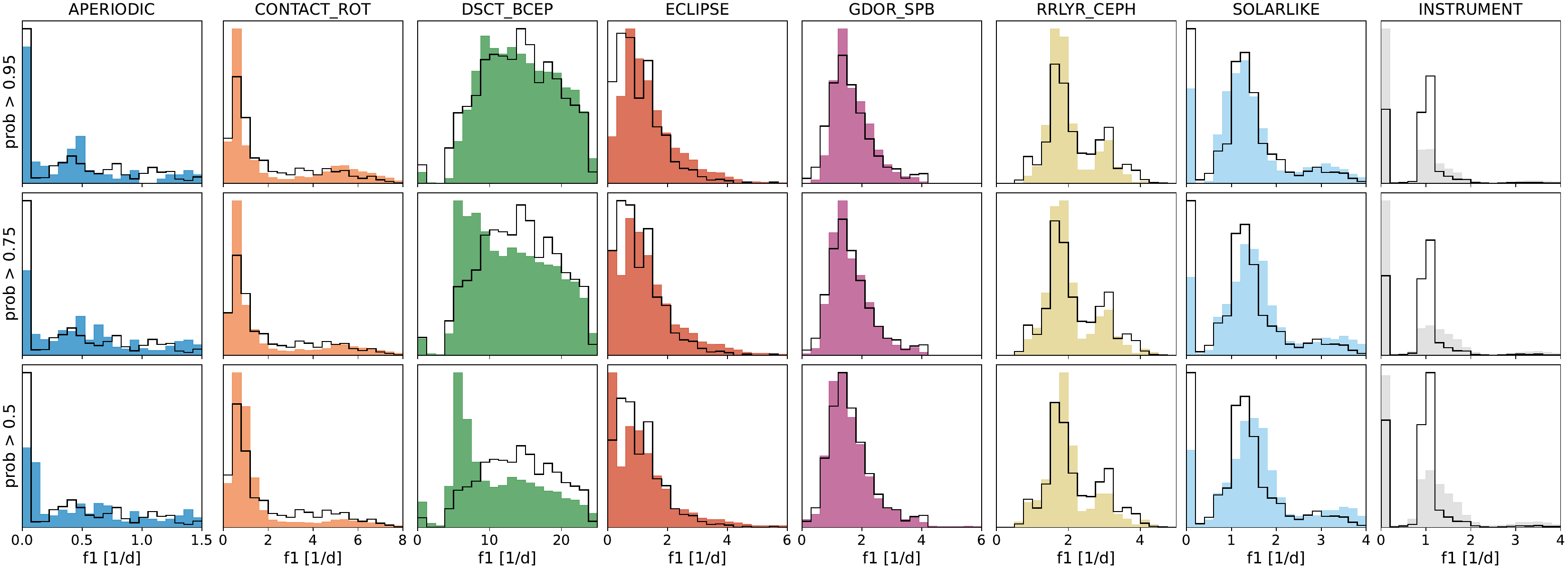}
    \caption{Normalised distributions of dominant frequencies $f_1$ for classified light curves (in colour) with final scores above 0.95 (top), 0.75 (middle), and 0.5 (bottom) and those of the training set (black outline). If a single \gaia DR3 had multiple light curves, all of them were included.}
    \label{fig:f1}
\end{figure*}

Seven out of the eight classes in our pipeline are variable stars. Inquiries can therefore be made for the basic variability properties of the unlabelled and labelled sets. Fig.\,\ref{fig:f1} shows the distributions of $f_1$ for stars that received 95\% (top row), 75\% (middle row), and 50\% (bottom row) probability for each of the eight classes from our pipeline (plotted in colour) compared to the training set distributions (in black). The distributions for all classes mostly follow the training set well for all score thresholds. One exception, which noticeably improves with the pseudo-probabilistic score increase, is the class of p-mode pulsators. Its distribution not only improves in shape, but also its height in the instrumental low-frequency bin fits better.

Fig.\,\ref{fig:a1} shows the accompanying distributions of $A_1$, which also follow closely the training set for all three chosen threshold values. A notable improvement occurs for the amplitude distribution for the RRLYR\_CEPH class, which gets closer to the one of the training set when the score increases. We achieved an overall significant improvement for this distribution compared to \citet[][]{gregory2026astrafier} following our injection of the training set with Cepheid stars to balance out RR Lyrae stars (Sect.\,\ref{sec:training}). Another improvement with the threshold increase is the shift of the SOLARLIKE distribution to lower values, as expected from that class in single-sector TESS light curves. This suggests that a 0.5 cut in scores might be too generous for these light curves, indicating possible confusions with g-mode pulsators or rotational variables. We finally note that the tails of the training set distributions are tighter than the ones for the unlabelled set, particularly for p- and g-mode pulsators and rotational variables, meaning that we are finding candidates with higher amplitudes.

\begin{figure}
    \centering
    \resizebox{\hsize}{!}{\includegraphics{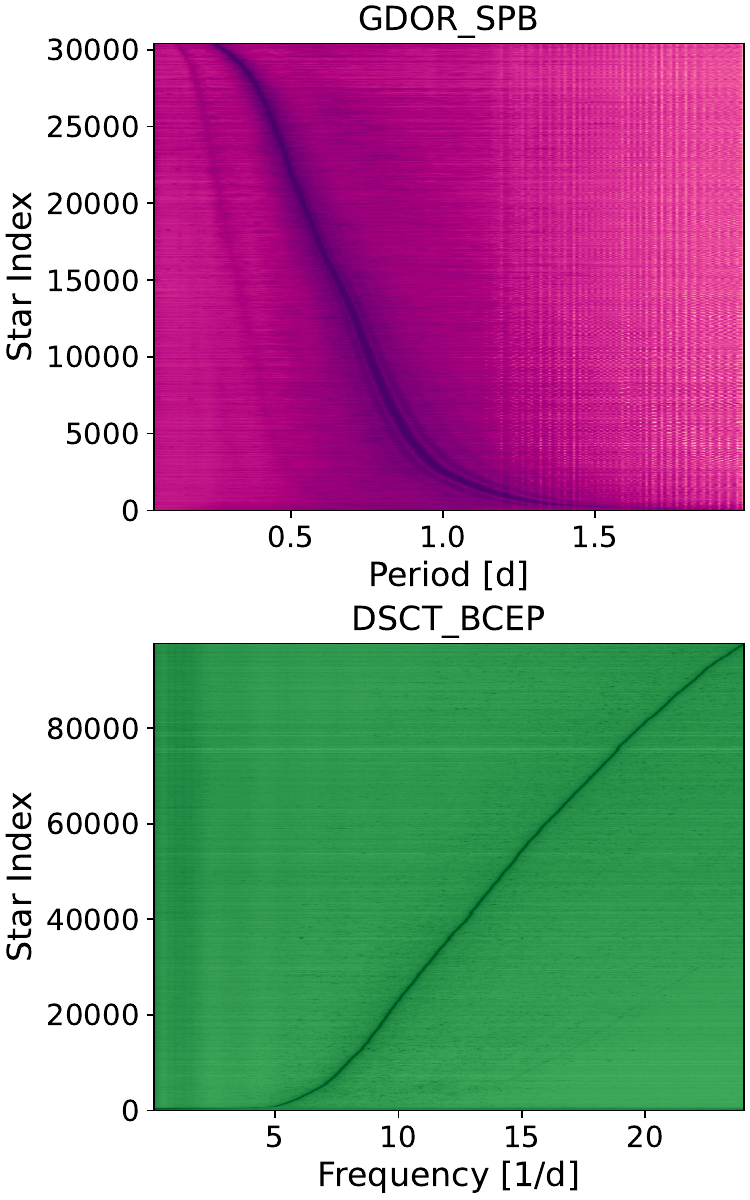}}
    \caption{Stacked amplitude spectra for candidate g-mode (in period, top panel) and p-mode (in frequency, bottom panel) pulsators. Only candidate light curves with final scores above 0.95 are plotted for visibility purposes.}
    \label{fig:stacked}
\end{figure}

We also constructed stacked amplitude spectra for g-mode (in period, top panel in Fig.\,\ref{fig:stacked}) and p-mode pulsators (in frequency, bottom panel in Fig.\,\ref{fig:stacked}) for classified unlabelled light curves. Each row in each subplot comprises an amplitude spectrum normalised against the height of the dominant peak (in logarithmic scale) in period (top) and frequency (bottom) space for an individual light curve, sorted by $1/f_1$ (top) and $f_1$ (bottom). The candidate p-mode pulsators in the bottom panel display a clean main ridge free of contaminants, similar to what was found in the study of $\beta\,$Cep stars by \citet[][]{fritzewski2025mode}. Additionally, a faint secondary ridge, which could be second-overtone p modes, is seen as in $\delta$ Sct stars found earlier in the literature \citep[][]{ziaali2019period,barac2022revisiting,hey2024confronting}. The first-overtone p modes are likely invisible due to the chosen normalisation along with the high number of stacked rows.

For the g-mode pulsators in the top panel, the main ridge is expected for pulsators with dominant 
dipole prograde $(l,m)$~$=$~$(1,1)$ modes \citep[][]{li2020gravity,aerts2024asteroseismic,hey2024confronting}. The lower $l=2$ mode ridge \citep[][]{li2020gravity,hey2024confronting} and the upper r-mode ridge \citep[][]{li2020gravity}, both with lower amplitudes than for $f_1$, can also be clearly seen. This shows that our classification pipeline properly finds candidate p- and g-mode pulsators
with expected properties and displaying multiperiodic pulsation behaviour, making them interesting targets for follow-up research, particularly when more than one type of mode is already observed in a single TESS sector.

\subsection{Rotational properties}

\begin{figure}
    \centering
    \resizebox{\hsize}{!}{\includegraphics{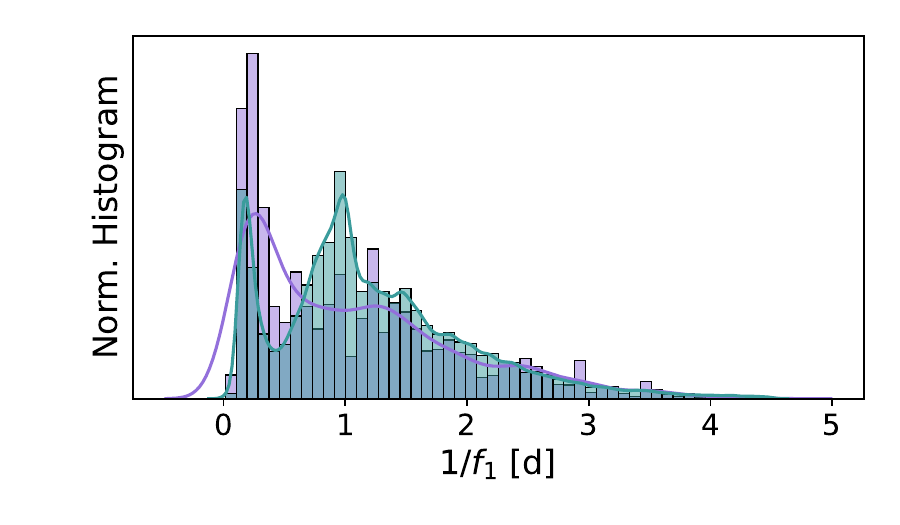}}
    \caption{Normalised distributions of the dominant period $1/f_1$ for the training set (purple) and classified light curves with final scores above 0.5 (teal). Due to small $1/f_1$ errors, both histograms and Kernel Density Estimators (as full lines) were computed directly from the point estimates.}
    \label{fig:prot}
\end{figure}

Aside from TESS's power for pulsational studies, also rotational variability also occurs prominently. In
Fig.~\ref{fig:prot} we show distributions of the dominant period $1/f_1$ for the CONTACT\_ROT class 
already shown in frequency in the second column in Fig.~\ref{fig:f1}. 
The distributions of $1/f_1$, which are either rotational periods or half rotational periods, are skewed towards lower values and indicate a preference for short-period rotational variables. This is consistent with \citet[][]{colman2024methods} and \citet[][]{hattori2025measuring}. While both distributions are bimodal, 
their relative heights differ, suggesting that the CONTACT\_ROT class tends to favour rotational variables over contact binaries in the predictions.

\begin{figure}
    \centering
    \resizebox{\hsize}{!}{\includegraphics{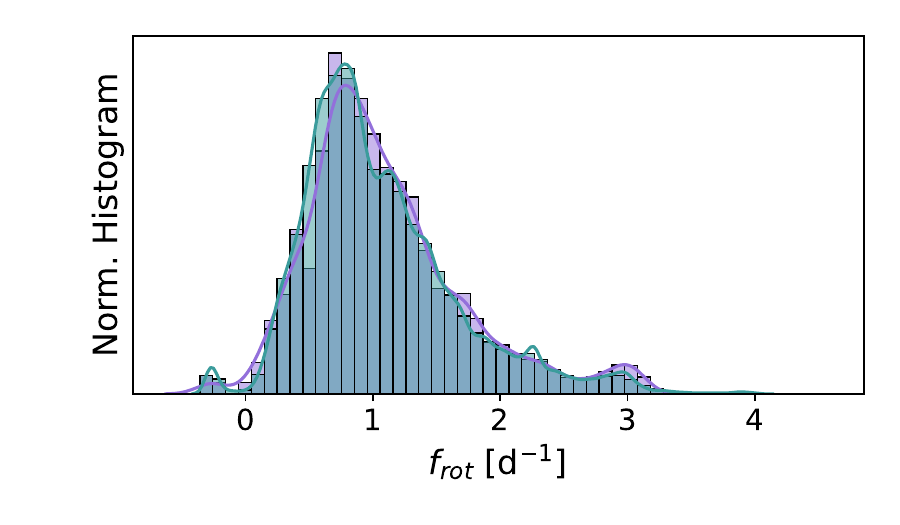}}
    \caption{Normalised distributions of the near-core rotation frequency $f_{\rm rot}$ computed from the recipe in \citet{aerts2025evolution} for the training set (purple) and for classified light curves with final scores above 0.5 (teal). Histograms are plotted from 1\,000 samples for each candidate frequency within the uncertainty range. Kernel Density Estimators are plotted as full lines and were computed directly from the point estimates.}
    \label{fig:frot}
\end{figure}

From the perspective of stellar rotation, the class of g-mode pulsators is of particular interest. Assuming that the dominant frequency $f_1$ for stars labelled as GDOR\_SPB is a dipole prograde mode, which is a sensible approach following the findings by \citet[][]{li2020gravity} and \citet[][]{Pedersen2021}, we estimated the near-core rotation frequency $f_{\rm rot}$ from the recipe deduced by \citet{aerts2025evolution}: 
$f_{\mathrm{\rm rot}} = 0.836^{+0.023}_{-0.027}\, f_{\mathrm{1}}
 - 0.272^{+0.041}_{-0.036}\ \mathrm{d}^{-1}$. We do note that this regression formula is only valid for limited ranges of $f_1$ (we refer to \citet[][]{aerts2025evolution} for details on the conditions and
asteroseismic applications). We show the distributions of $f_\mathrm{\rm rot}$ for our sample in Fig.~\ref{fig:frot}, both for the training (purple) and unlabelled (teal) sets. They are mostly in line with 
observed near-core rotation rates of intermediate-mass stars \citep[][]{Aerts2025-rot} but our study added fast-rotating pulsator candidates. These targets are of high interest for follow-up studies, particularly for  future asteroseismic modelling of stars flattened by the centrifugal force \citep[cf.\,][for a review of asteroseismology of fast rotators]{aerts2024asteroseismic}.

Finally, based on a manual inspection of light curves, we found that both the training set of p-mode pulsators and the stars labelled as DSCT\_BCEP display rich pulsation spectra with rotationally-split multiples. Both candidate dipole ($l=1$) and quadrupole ($l=2$)
multiplets are found, some symmetric and other with asymmetry. Identification of these multiplets
in the unlabelled set as in \citet{Burssens2023, vanlaer2025interior, fritzewski2025mode}; Vanrespaille et al. (2026, in review),
will allow to measure the interior rotation frequency, as a promising future direction for catalogue users.

\subsection{LOPS2 variability in the HRD}\label{sec:hrd}

Aside from properties directly inferred from the TESS data, it is also meaningful to consider the stars' global parameters, e.g.\ by placing them into the HRD from a homogeneous treatment.
Even if the \gaia data for the effective temperature and luminosity can come with major (and often unknown) uncertainties, relative positions in the HRD are still helpful for variability classifications \citep[][]{de2023gaia}, particularly for p-mode \citep[e.g.,][]{Murphy2019} and g-mode pulsators \citep[e.g.,][]{Aerts2023}.
In Fig.\,\ref{fig:teff} we show the distributions of $T_{\mathrm{eff}}$ from \gaia DR3 (GSP-Phot) for stars that received 95\% (top row), 75\% (middle row), and 50\% (bottom row) probability for each of the eight classes from our pipeline (in colour) compared to the training set distributions (in black). For some classes, the distributions are shifted towards cooler temperatures, which on a case-by-case basis is either a genuine misclassification, an incorrect \gaia temperature, or a result of light curve contamination. Increasing the threshold improves the ratio of two modes in the distribution for main-sequence p- and g-mode pulsators, where the second mode (hotter stars) is noticeably higher for the 95\% (top row) threshold. However, overall we advise against using $T_{\mathrm{eff}}$ as a single diagnostic and/or filtering metric, and instead suggest combining it with other filters. We also note that the shape of the $T_{\mathrm{eff}}$ for APERIODIC class in the top row occurs because this class generally has lower probabilities than other classes. Hence fewer sources pass the score filtering.

\begin{figure}
    \centering
    \resizebox{\hsize}{!}{\includegraphics{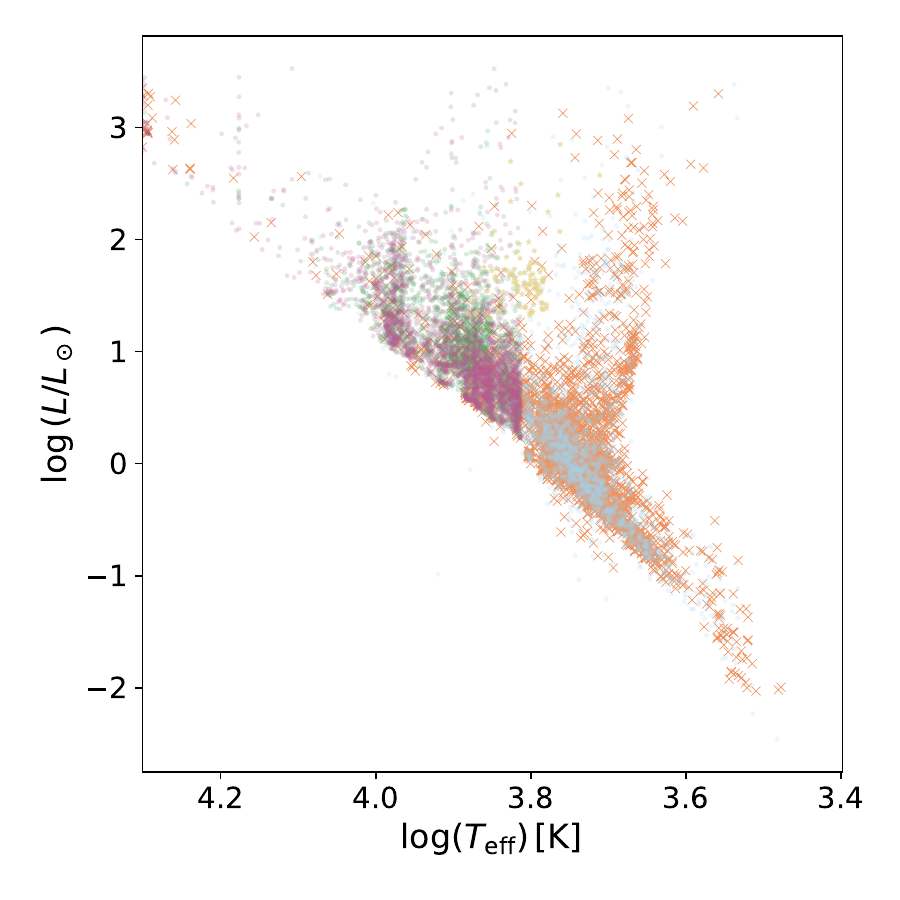}}
    \caption{HRD with a sub-sample of light curves classified as CONTACT\_ROT (orange crosses), DSCT\_BCEP (green), GDOR\_SPB (magenta), RRLYR\_CEPH (yellow stars), and SOLARLIKE (light blue). Vertical ridges are due to \gaia DR3 temperature grid systematics.}
    \label{fig:hrd}
\end{figure}

We applied a filtering recipe from Sec. \ref{sec:probas} to remove potential contaminants, most of which are attributed to blending. For the remaining sources, we computed the luminosities, $L$, from extinction-corrected absolute magnitudes and applied bolometric corrections to place the stars in the HRD, as displayed in Fig.~\ref{fig:hrd}. For visibility purposes, only 10\,000 of stars are plotted, with no duplicates per \gaia DR3 ID. This Fig. reveals
that high-probability pulsating stars are mostly located in instability regions expected for these types of pulsators \citep[][]{aerts2010asteroseismology,aerts2021probing}. Fig.~\ref{fig:hrd3} shows the position in the HRD for the remaining three classes (APERIODIC, ECLIPSE, INSTRUMENT), revealing that these stars are found all over the HRD, as expected.

As briefly mentioned in Sect.\,\ref{sec:ens}, one of the main advantages of a probabilistic multi-class machine learning pipeline is the potential to flag hybrid classes. For example, hybrid g- and p-mode pulsators are expected to receive higher probability for the class that captures its dominant variability and a lower but still significant score for the class of its secondary variability following the properties of such stars \citep{dupret2004,grigahcene2010,bowman2017,li2020gravity,audenaert2022,skarka2022,skarka2024,kliapets2025automated}.
We investigated this by plotting over unique 1\,000 targets in an HRD for targets with one assigned class probability $\ge 50\%$ and the other $\ge 20\%$ (top panel of Fig. \ref{fig:hrd2}). The light curve of such a source in the bottom panel of the Fig. clearly displays low- and high-frequency variability, making it a candidate hybrid pulsator.

\subsection{Some variable types beyond the eight classes}\label{sec:beyond}

The `variability tree' introduced by \citet[][]{Eyer2008} has many more types than the classes we considered here. Having discussed the behaviour of the eight trained classes, we thus checked the outputs for several other classes not included in our framework. Pulsating hot subdwarf (sdB) stars are faint blue compact objects in the core helium burning stage of their evolution
\citep[][for a comprehensive review]{Heber2009}.
A fraction of them have p~modes with periods of a few minutes and others reveal
g~modes of about an hour, with some hybrids having both \citep[][]{aerts2010asteroseismology}. Despite these short timescales posing a challenge for discovery from 30-min cadence TESS data, hundreds of candidates have been found in such data \citep[][]{Sahoo2020,Sahoo2023,Baran2021}. We checked where these stars end up in our catalogue. Most of them
are predicted as belonging to the INSTRUMENT class. Less than 4\% are classified as p- or g-mode pulsators or eclipsing binaries, each. This makes it unlikely to find these compact objects 
in our catalogue 
from filtering on scores only. However, thresholding on fundamental parameters and position on the colour-magnitude diagram should allow the user to identify candidate sdB stars \citep[][]{ranaivomanana2025variability,uzundag2025observing} in our variability catalogue, or make it possible to find them in future versions if an additional short-period pulsator class gets incorporated in the classification.

Light curves for the broad category of magnetic pulsators \citep[][]{thomson2025discovery} are mostly (57\%) predicted as p-mode pulsators, with the rest being labelled as g-mode pulsators, rotational variables, and several as INSTRUMENT. These predictions are consistent with sources queried, which included stars manually classified as $\delta$ Sct, $\gamma$ Dor, $\beta$ Cep, and SPB stars (Thomson-Paressant et al. 2026, in prep). This demonstrates that our ensemble approach is well capable of finding these stars and placing them in sensible respective (pulsator) classes. However, efficiently finding these sources on an industrial scale would likely require specialised features in a feature-based machine learning architecture. Given 
the high potential of magnetic pulsators in different mass regimes \citep[][]{thomson2020search,lecoanet2022asteroseismic,vandersnickt2025asteroseismic,takata2026asteroseismic}, this type of pulsators might be considered for future versions of the variability catalogue, provided a sufficient training sample can be gathered.

\section{Discussion on classifier design and performance}\label{sec:disc}

This study provided 
a potential benchmark dataset for all-sky variability studies not connected to PLATO. Despite high performance in terms of classification results and parameter space covered, this study revealed several limitations and room for future improvements stemming from used methods, data, and instruments before we upgrade it to an all-sky variability assessment and catalogue.

From the methodological perspective, the deep learning-based classifier is an experimental development. Application of deep learning in astronomy is a fast developing field \citep[][for a review]{Ting2025}, with best practices and optimal framework configurations still being discovered. The particular neural network 
we used from \citet[][]{gregory2026astrafier} was designed to identify similarities based on long- and short-term variability, and overall light curve morphology. Neural networks in general are particularly good at recognising data that closely resembles the training set, but occasionally work in ways other than intended by the developer and/or user. We showed an example of this from our
manual inspection of light curves that received high scores without exhibiting stellar variability
as the source in Fig.~\ref{fig:ens_err}.

Confusion between instrumental effects and genuine variability mostly affects the class of p-mode pulsators. This may have various reasons. First, the training set might not be optimally representative for
all possible light curve systematics that appear in the TESS instrument and/or the TGLC pipeline. The training set comes from a limited number of TESS sectors (and we chose to downgrade to 30-min cadence), while the predictions are made on a much larger sector selection with each of different data quality. Second, transformers receive normalised data, as part of the neural network architecture. This means that information about the amplitude of the stellar signal is not used by the model. Finally, a neural network might be taking a neural shortcut \citep[][]{geirhos2020shortcut}, meaning it could sometimes be relying on non-generalisable features instead of looking at actual variability. While neural nets do offer an advantage of not having to engineer and select features, these potential caveats matter because of the overly optimistic predictions, prioritising high scores over split pseudo-probabilities. This may lead to misinterpretations by the end-user \citep[][]{guo2017calibration}, as we chose not to calibrate predictions for consistency with XGBoost.
Our ensemble approach partially mitigates these caveats, delivering largely robust stellar variability classifications, especially for the cases where light curves are not overly contaminated with instrumental effects.

Gradient-boosting trees, on the other hand, do require feature engineering, which is a process that relies on substantial domain knowledge and is often specific to a particular learning problem. Decision trees used in the ensemble rely on a limited set of features, which do not exhaust the rich realm of features that can be used for variability classification. Recent variability studies \citep[][]{huijse2025learning,ranaivomanana2025variability} demonstrated that \gaia can offer a treasure trove of information for automated variable star research. One possible future improvement includes incorporating differences in temperature (or colour as it is model-independent) from \gaia into the feature space. Among other things, it would allow the division of main-sequence pulsators into separate categories (e.g.\ $\gamma$ Dor and SPB stars; \citealt{hey2024confronting}), although non-radial pulsators as seen by \gaia DR3 seem to constitute one uninterrupted continuum along the main sequence
\citep[][]{de2023gaia,Mombarg2024,aerts2025evolution}.

From the data perspective, in addition to splitting p- and g-mode pulsators into sub-classes via features, similar exercises can be done for classes of rotational variables and contact binaries. Such work could be done in parallel with enriching the training set with examples of these classes from other TESS sectors. Additionally, more classes, such as pulsating eclipsing binaries \citep[][]{kemp2025populations}, hump-and-spike \citep[][]{saio2018theory,antoci2025magnetic}, or heartbeat stars \citep[][]{welsh2011koi,hambleton2013physics,li2024twenty}, in addition to the ones already mentioned in Sect.\,\ref{sec:beyond}, could become separate classes. We refrained from such generalisations due to lack of sufficient carefully selected validated training examples from current TESS photometry.

\section{Conclusions}\label{sec:conc}

In this work, we classified 38 million TESS light curves in PLATO's LOPS2. This resulted in the first version of the open-source variability catalogue, which may help guide the selection of targets for future PLATO GO studies. We used two different machine learning models and combined their prediction into an ensemble to reduce prediction variance, resulting in a more limited but purer, high-score sub-sample of 3.59 million stars (unique sources) belonging to variable -- non-instrumental -- classes.

In addition to the identifying information (TIC, \gaia DR3 IDs, and coordinates) and scores for each of the eight variable classes, the variability catalogue includes value-added information (temperature, colour, luminosity, $G$-band magnitude, TESS flux contamination index, the number of PLATO N-CAMs pointing at the target, and whether the two classifiers agree on the highest label assignment) to help make informed selections for 
follow-up studies, including target selections for future PLATO GO proposals.

Potential pipeline improvements in the short term include refining and expanding the training set, adding more variable classes, engineering features tailored to identify specific types of (pulsating) stars, and developing web-interfaces to access and navigate the variability catalogue. 
A specific future refinement of particular potential for asteroseismology of g-mode pulsators is to include many more modes for the classes of multiperiodic non-radial pulsators from reliably estimating their frequencies and amplitudes (see e.g. \citealt{Scott2026}). Stitching light curves from multiple sectors is a promising avenue for such future research, particularly considering a partial overlap of LOPS2 with a Continuous Viewing Zone of TESS \citep[][]{ricker2015transiting,nascimbeni2025plato}. However, this presents challenges of using data with inconsistent quality from various TESS sectors, which might require a new approach to data preprocessing. Additionally, this would result in using different lengths light curves, which requires adaptation in the methodology. 

A caveat is that the knowledge transfer of labels from one instrument to another is challenging. Therefore, it is important to remember that the variability catalogue is a classification of light curves, not necessarily objects. For example, a hybrid pulsator in \kepler might appear g-mode dominated, while SPOC, QLP, TGLC aperture, and TGLC PSF photometry may each capture the same signal differently \citep[][]{kliapets2025automated}. As a result, the same object may appear g-mode dominated in some light curves and p-mode dominated in others. A similar situation is expected from PLATO light curves, further complicated by different noise budgets for different combinations of N-CAMs.

In the longer term, future work to improve the classification can focus on various aspects, such as including information from other surveys like \gaia \citep[][]{prusti2016gaia}, SDSS-V \citep[][]{kollmeier2026sloan}, or PLATO \citep[][]{rauer2025plato} to create multimodal machine learning frameworks \citep[][]{huijse2025learning,rizhko2025astrom3}. 
Once the first PLATO data are available, they can be integrated into the catalogue pipeline. A particularly interesting prospective application is the use of foundational models to disentangle true variable from telescope-specific instrumental signal  \citep[][]{Audenaert2025b,Mercader2026}, which can be trained on multimodal datasets such as the Multimodal Universe \citep[][]{angeloudi2024multimodal}.

\section*{Data availability}

Files uploaded on Zenodo include: (i) the training set; (ii) the variability catalogue; and (iii) the models.

\begin{acknowledgements} 
    We are grateful to Victoria Antoci and Joris De Ridder for useful discussions \& René Heller, Beatriz Mas Sanz, and Tiziano Zingales for comments related to this paper. MK and CA acknowledge The Kavli Foundation for their financial support in the framework of the Kavli Scholarship given to MK from 25/9/2023-24/9/2025, including facilitation of MK's research visit to the MIT Kavli Institute for Astrophysics and Space Research in the fall of 2025 (hosts: JA and GRR), which ultimately led to the current paper. CA, PH, MV, PA, and RSS acknowledge financial support from the European Research Council (ERC) under the Horizon Europe programme (Synergy Grant agreement N°101071505: 4D-STAR). While partially funded by the European Union, views and opinions expressed are however those of the author(s) only and do not necessarily reflect those of the European Union or the European Research Council. Neither the European Union nor the granting authority can be held responsible for them. CA, DJF, JV, HW, ROA, and RSS acknowledge financial support from the Flemish Government under the long-term structural Methusalem funding program by means of the project SOUL: Stellar evOlution in fUll gLory, grant METH/24/012 at KU Leuven. 
    AT, PH, NJ, and CA are thankful for the funding by the BELgian federal Science Policy Office (BELSPO) through PRODEX grants PLATO and {\it Gaia}. 
    MK, PH, DF, and CA receive funding from the Research Foundation Flanders (FWO) under grant G051626N.
    Some funding for this project was supplied by MIT’s Undergraduate Research Opportunities Program (UROP).
    DMB gratefully acknowledges UK Research and Innovation (UKRI) in the form of a Frontier Research grant under the UK government's ERC Horizon Europe funding guarantee (SYMPHONY; PI Bowman; grant number: EP/Y031059/1), and a Royal Society University Research Fellowship (PI Bowman; grant number: URF{\textbackslash}R1{\textbackslash}231631). SJM was supported by the Australian Research Council through Future Fellowship FT210100485. NS acknowledges financial support from the project “Young Stellar Objects observed by PLATO: sample characterisation and automatic classification,” (FFG project number: FO999928671), funded by the FFG (www.ffg.at). GL is being supported by the Australian Research Council through the DECRA project DE250100773. VV gratefully acknowledges support from the Research Foundation Flanders (FWO) under grant agreement N◦1156923N (PhD Fellowship). This work was supported by Fundação para a Ciência e a Tecnologia (FCT) through national funds under grant UID/04434/2025 and project 2022.03993.PTDC (DOI: 10.54499/2022.03993.PTDC); I.R. is funded by i) FCT through grant 2024.01287.BD, ii) Bolsas da Caixa Geral de Depósitos (CGD), and iii) project MIND PhD24 through grant MP25PHD0136. JMB and EP were partially supported by the `SeismoLab' KKP-137523 Élvonal grant
of the Hungarian Research, Development and Innovation Office (NKFIH).
    This work presents results from the European Space Agency (ESA) space mission PLATO. The PLATO payload, the PLATO Ground Segment and PLATO data processing are joint developments of ESA and the PLATO mission consortium (PMC). Funding for the PMC is provided at national levels, in particular by countries participating in the PLATO Multilateral Agreement (Austria, Belgium, Czech Republic, Denmark, France, Germany, Italy, Netherlands, Portugal, Spain, Sweden, Switzerland, Norway, and United Kingdom) and institutions from Brazil. Members of the PLATO Consortium can be found at https://platomission.com/. The ESA PLATO mission website is https://www.cosmos.esa.int/plato. We thank the teams working for PLATO for all their work. This paper includes data collected by the TESS mission. Funding for the TESS mission is provided by the NASA’s Science Mission Directorate. This work has made use of data from the European Space Agency (ESA) mission Gaia (https://www.cosmos.esa.int/gaia), processed by the Gaia Data Processing and Analysis Consortium (DPAC, https://www.cosmos.esa.int/web/gaia/dpac/consortium). Funding for the DPAC has been provided by national institutions, in particular the institutions participating in the Gaia Multilateral Agreement.
\end{acknowledgements}

\bibliographystyle{aa}
\bibliography{bib.bib}

\onecolumn
\begin{appendix}

\section{Dataset noise budget}

\begin{figure}[htbp]
    \centering
    \includegraphics[width=18cm]{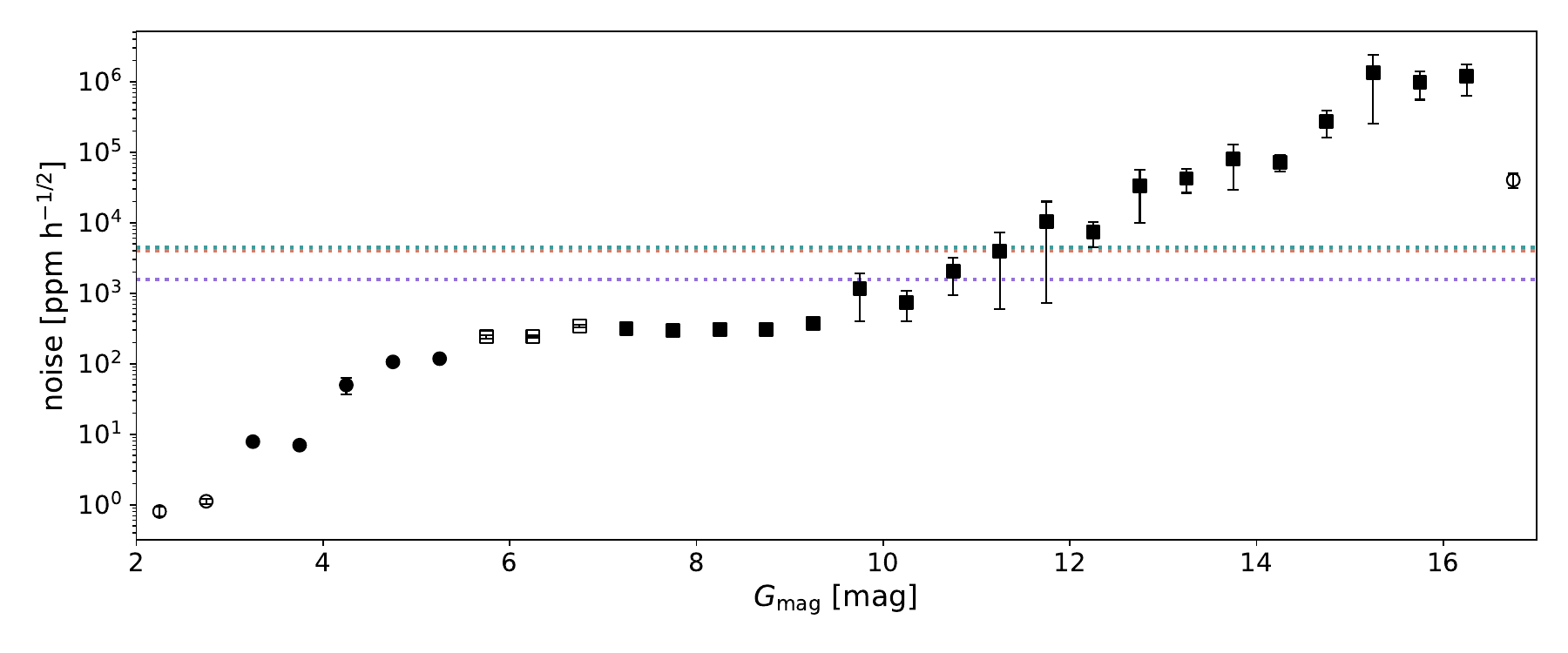}
    \caption{Noise budget -- with standard errors -- of unlabelled TESS light curves as a function of \gaia magnitude, binned in magnitude increments of 0.5. Empty circles, full circles, empty rectangles, and full rectangles indicate total number of light curves in each $G_{mag}$ bin in the catalogue: $\le100$, $\le1\,000$, $\le5\,000$, and $\ge 5\,001$, respectively. We sampled $10\,000$ light curves from each bin if they were available; if less were available, all of them were used. Horizontal dotted lines indicate average noise budgets of the extended-mission (purple), downsampled extended-mission (teal), and nominal-mission (burgundy) data from the training set; standard errors have been omitted for visibility.}
    \label{fig:noise}
\end{figure}

\section{Fraction of predicted classes per magnitude}

\begin{figure}[htbp]
    \centering
    \includegraphics[width=18cm]{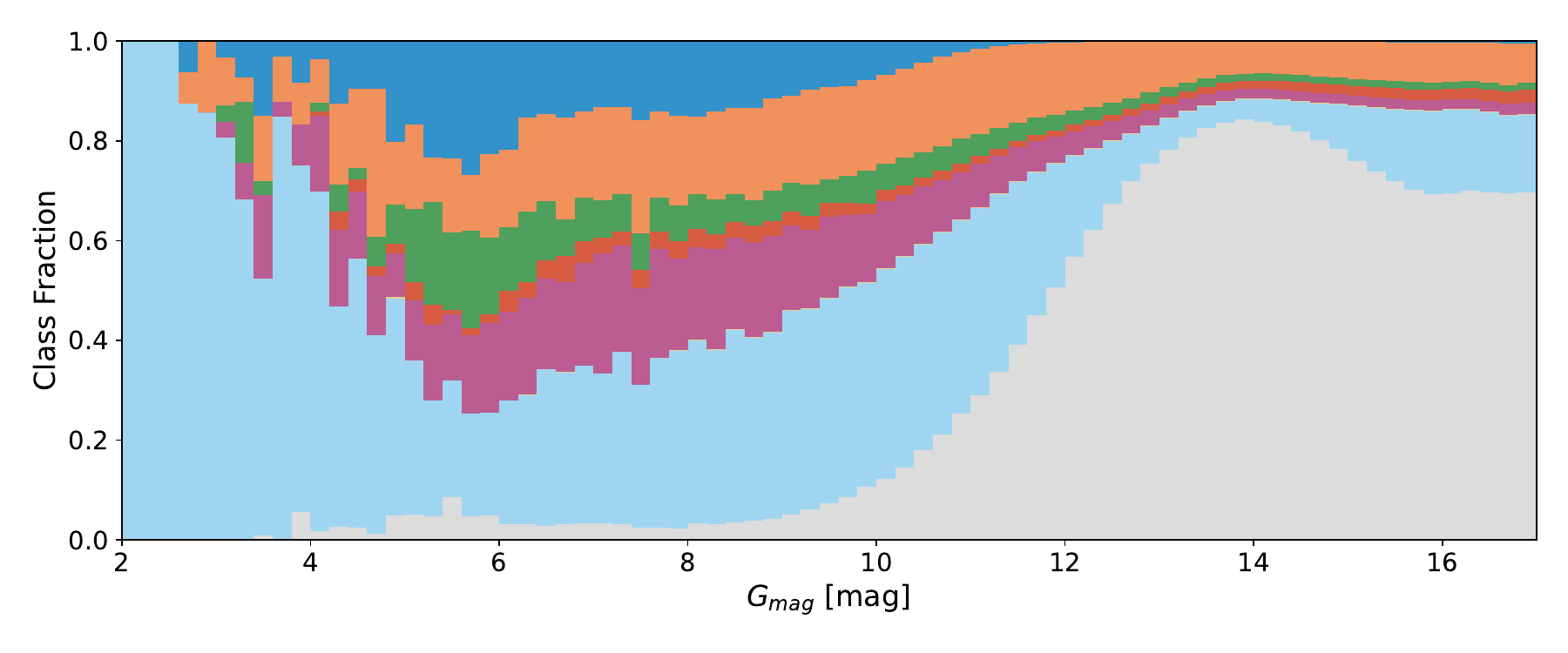}
    \caption{Fraction of light curves predicted for each class -- APERIODIC (dark blue), CONTACT\_ROT (orange), DSCT\_BCEP (green), ECLIPSE (red), GDOR\_SPB (magenta), RRLYR\_CEPH (yellow), SOLARLIKE (light blue), INSTRUMENT (grey) -- as a function of \gaia magnitude.}
    \label{fig:frac}
\end{figure}

\clearpage

\section{Temporal stability case study}\label{appendix:downs}

\begin{figure}[htbp]
    \centering
    \includegraphics[width=18cm]{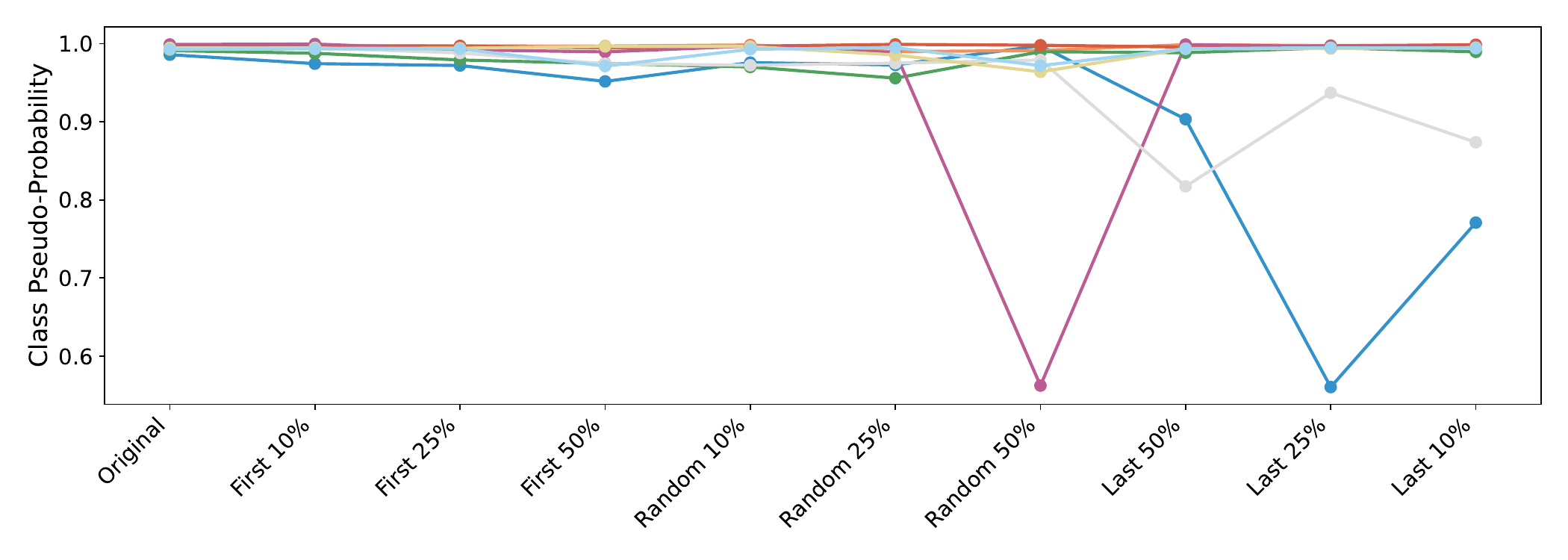}
    \caption{Predictions for sub-sampled light curves of each class: APERIODIC (dark blue), CONTACT\_ROT (orange), DSCT\_BCEP (green), ECLIPSE (red), GDOR\_SPB (magenta), RRLYR\_CEPH (yellow), SOLARLIKE (light blue), INSTRUMENT (grey). The first tick on the x-axis represents the original light curve prediction. Each subsequent tick represents a sub-sampled light curve, e.g.\ "First 10\%" means that the first 10\% of the flux and time points have been removed.}
    \label{fig:permut}
\end{figure}

We evaluated temporal stability of our ensemble by choosing a random time series with a final probability for one of the eight classes above 0.99 (0.98 for APERIODIC as no light curve received a final score for that class $> 0.99$). We then sub-sampled and predicted on it after removing the respective flux and time points. 
More concretely, we remove 10\%, 25\%, and 50\% of the points from the beginning of the time series, 10\%, 25\%, and 50\% of the points from the end of the time series, and 10\%, 25\%, and 50\% of the points chosen randomly. Results of predicting on these sub-sampled time series are shown on Fig.\,\ref{fig:permut}.

The most adversely-affected case occurs for a candidate g-mode pulsator, for which the score decreased by 43.67\%. The light curve for an APERIODIC candidate is affected mostly by removing the last \{10, 25, 50\}\% datapoints. The other variable classes were largely unaffected. Notably, none of the permutations changed the highest class assignment for any of the eight chosen light curves. This suggests that our ensemble approach is temporally-stable and can be applied to time series shorter than an average TESS sector length.

\section{Consistency across sectors}\label{appendix:consist}

\begin{wrapfigure}{r}{0.45\textwidth}
    \centering
    \includegraphics[width=0.43\textwidth]{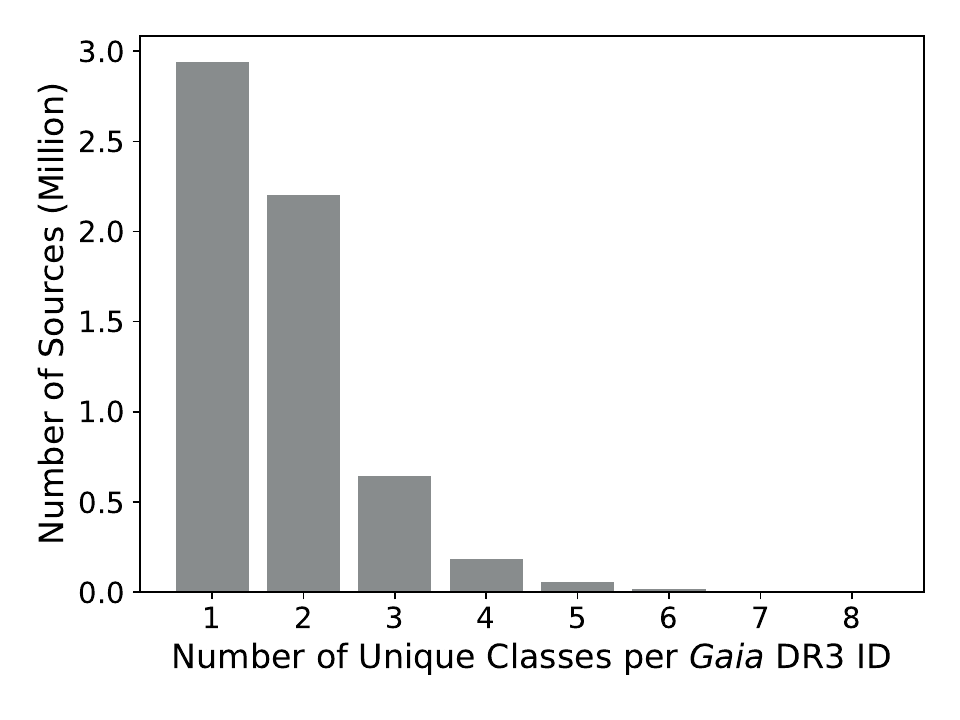}
    \caption{Distribution of assigned classes per target according to its \gaia DR3 ID.}
    \label{fig:mult}
\end{wrapfigure}

Given that most targets in LOPS2 have been observed in multiple TESS sectors, we checked how many objects receive different predictions sector-to-sector. This may occur because each TESS sector can have its own instrumental effects influencing the morphology of the light curve, such as amplitudes of modes in hybrid classes. Sector-to-sector changes may also occur when the two used models agree on some sector classifications but not on another.

Fig.~\ref{fig:mult} shows the distribution of how many different best-class assignments 
occur in the unlabelled dataset per \gaia DR3 ID across sectors. Most sources only have one class assigned to them. Among the sources that have two classes assigned, the most common is the instrumental and solar-like pair, followed by the pairs of low-frequency classes, that is, rotational variables assigned to the INSTRUMENT, SOLARLIKE, and GDOR\_SPB classes. The fifth highest pair is the combination of p-mode pulsators with the instrumental class, which is further discussed in Sect.\,\ref{sec:disc}. Combinations beyond two classes drop with the increase in the number of classes assigned, indicating overall class consistency. Only 1.5\% of the sources display more than four different high-confidence predictions from different TESS sectors.

\clearpage

\section{Example of blending}

\begin{figure}[htbp]
    \centering
    \includegraphics[width=16cm]{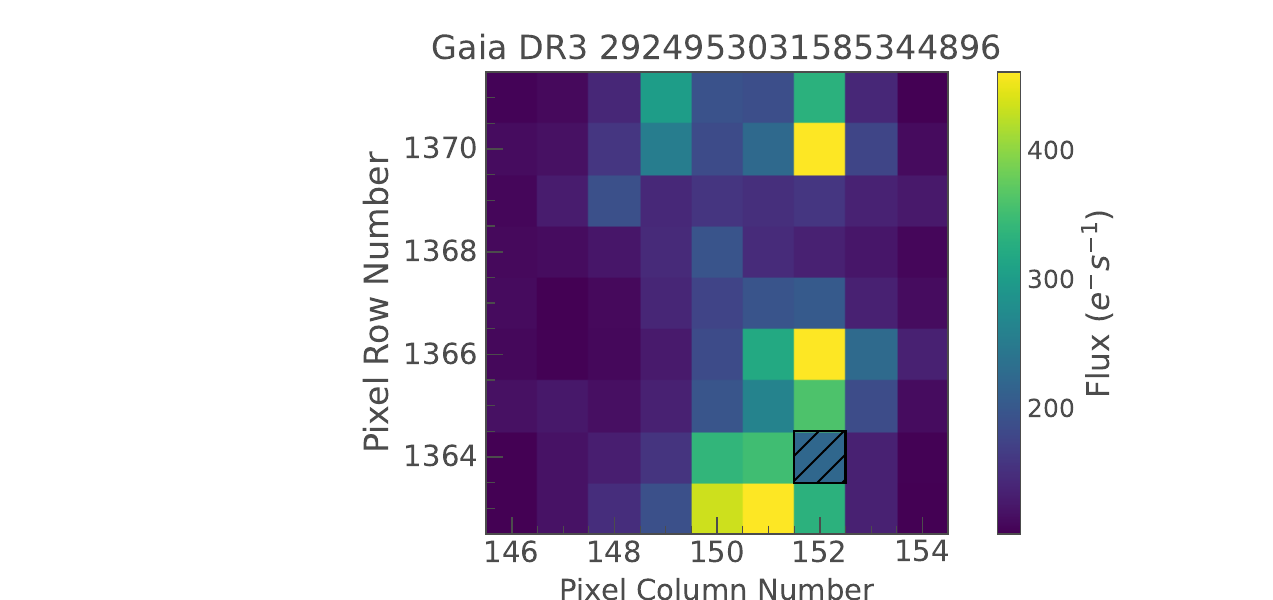}
    \caption{An example of a blended source (masked), \gaia DR3 2924953031585344896 (sector 33), contaminated by a bright neighbouring RR\,Lyrae star.}
    \label{fig:a3}
\end{figure}

\clearpage

\section{Distributions of highest amplitudes and \gaia DR3 effective temperatures}

\begin{figure}[htbp]
    \centering
    \includegraphics[width=18cm]{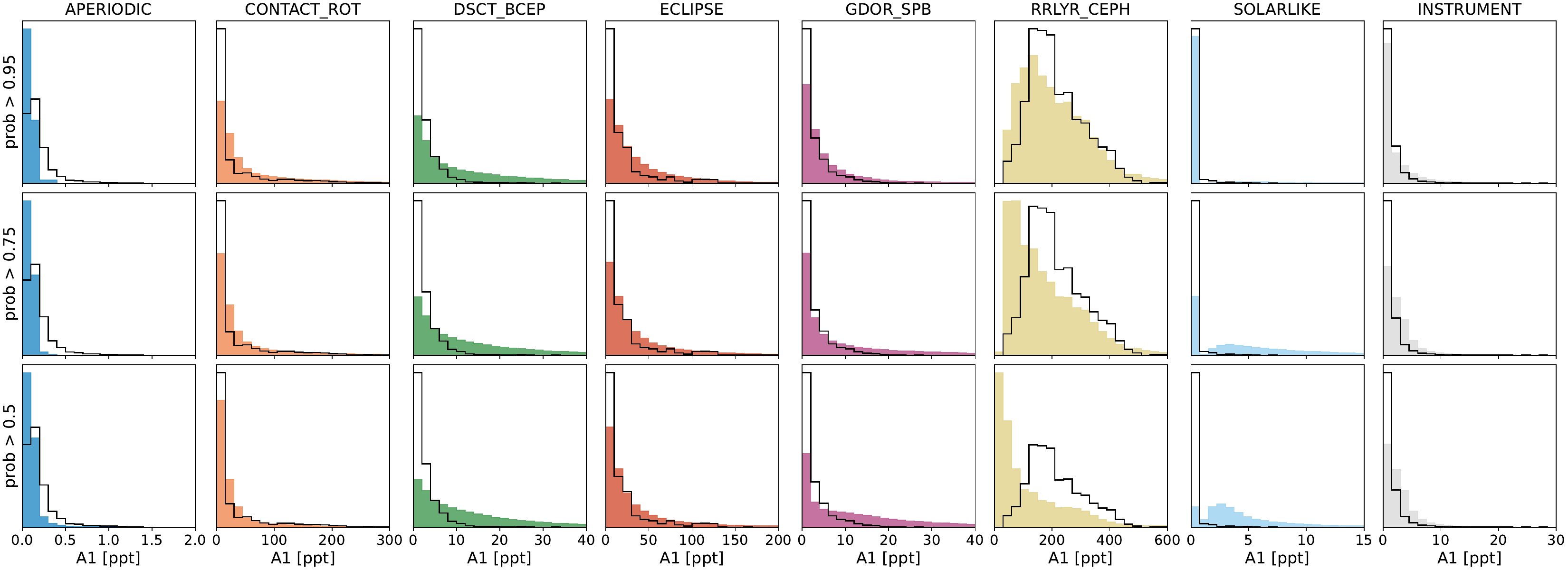}
    \caption{Same as Fig.\,\protect\ref{fig:f1} but for the highest amplitudes $A_1$.}
    \label{fig:a1}
\end{figure}

\begin{figure}[htbp]
    \centering
    \includegraphics[width=18cm]{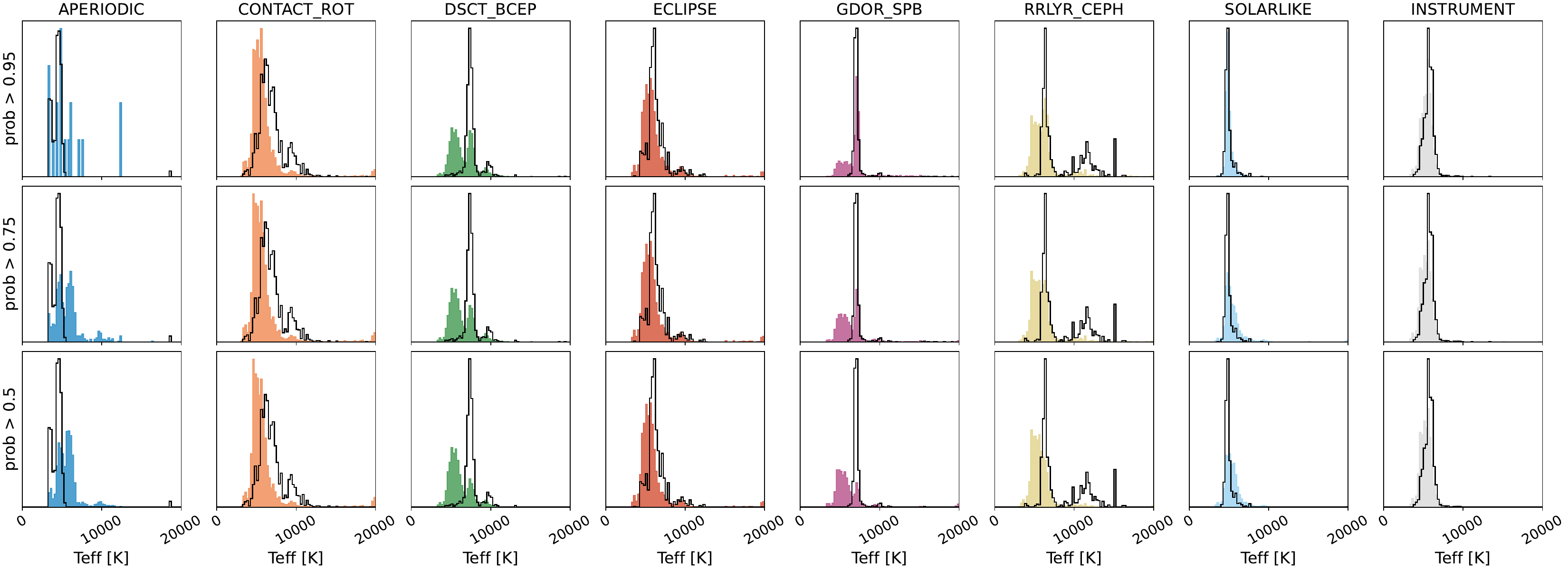}
    \caption{Same as Fig.\,\protect\ref{fig:f1} but for the distributions of effective temperatures $T_{\mathrm{eff}}$ from \gaia DR3.}
    \label{fig:teff}
\end{figure}

\clearpage

\section{HRD for remaining and hybrid classes}

\begin{figure}[htbp]
    \centering

    \begin{minipage}{0.48\textwidth}
        \centering
        \includegraphics[width=\linewidth]{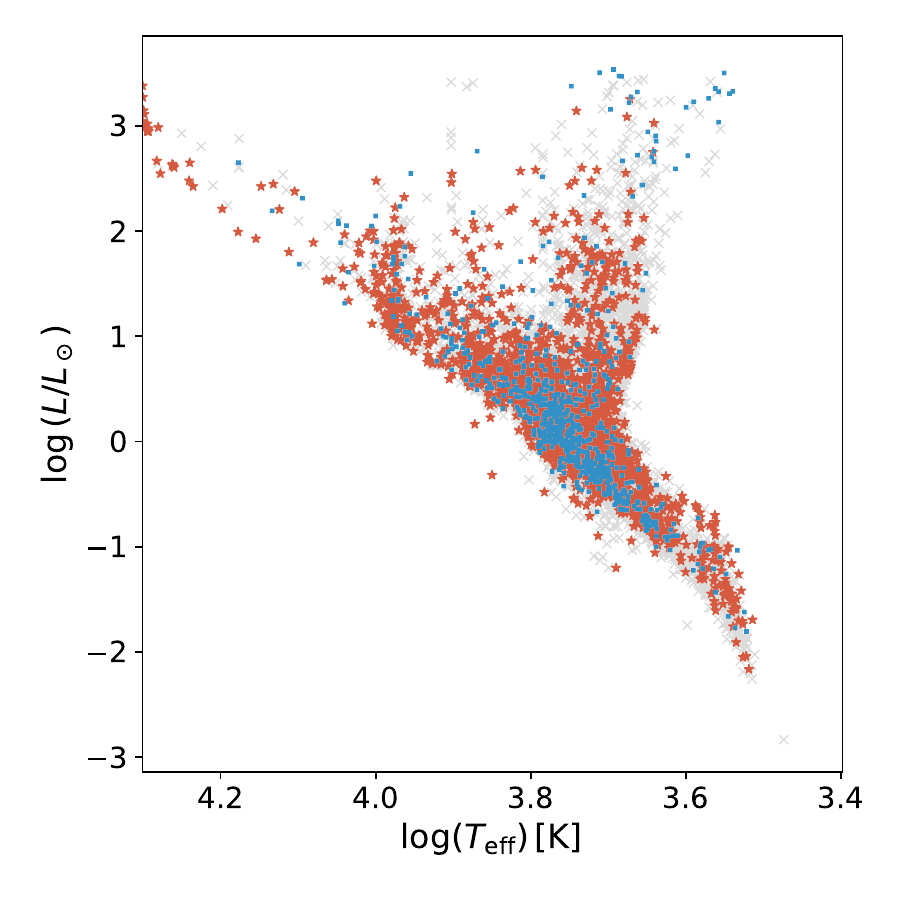}
        \caption{HRD with a sub-sample of light curves classified as APERIODIC (dark blue squares), ECLIPSE (red stars), and INSTRUMENT (grey crosses). For visibility, only a sub-sample of 15\,000 sources is shown.}
        \label{fig:hrd3}
    \end{minipage}
    \hfill
    \begin{minipage}{0.48\textwidth}
        \centering
        \includegraphics[width=\linewidth]{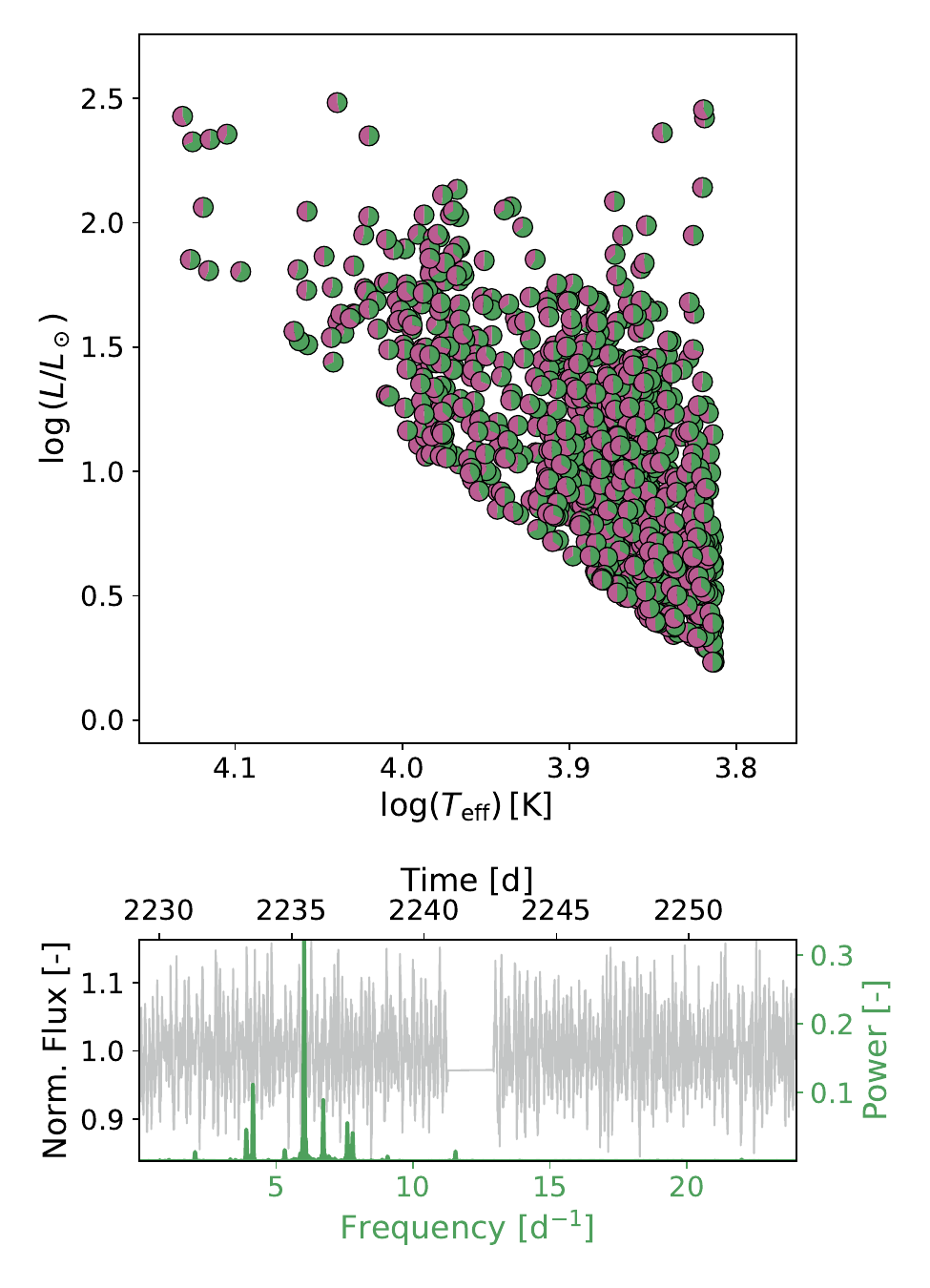}
        \caption{Top panel: HRD with all filtered sources classified as DSCT\_BCEP (green) or GDOR\_SPB (magenta), with pseudo-probabilities for one above 0.5 and for another above 0.2. Each point is a normalised pie chart of the probability distribution for just these two classes. 
        Bottom panel: same as Fig.\,\protect\ref{fig:ens_err} for \gaia DR3 5617518469549766144 (TESS sector 34), classified as DSCT\_BCEP.}
        \label{fig:hrd2}
    \end{minipage}

\end{figure}

\end{appendix}

\end{document}